\documentclass[preprint,12pt]{elsarticle}
\usepackage{amssymb}
\usepackage{graphicx}
\usepackage[utf8]{inputenc}
\usepackage{amsmath}
\usepackage{amsthm}
\usepackage{amssymb}
\usepackage[colorlinks=true]{hyperref}
\usepackage{subfig}
\usepackage[left=2.5cm,right=2.5cm]{geometry}
\usepackage[english]{babel}
\usepackage[toc,page]{appendix} 
\usepackage{tabularx} 

\usepackage{tikz}
\usetikzlibrary{shapes,arrows}

\newcounter{bla}
\newenvironment{refnummer}{%
\list{[\arabic{bla}]}%
{\usecounter{bla}%
 \setlength{\itemindent}{0pt}%
 \setlength{\topsep}{0pt}%
 \setlength{\itemsep}{0pt}%
 \setlength{\labelsep}{2pt}%
 \setlength{\listparindent}{0pt}%
 \settowidth{\labelwidth}{[9]}%
 \setlength{\leftmargin}{\labelwidth}%
 \addtolength{\leftmargin}{\labelsep}%
 \setlength{\rightmargin}{0pt}}}
 {\endlist}
\newcommand{\n }{\\ \nonumber }
\newcommand{\Thet }{\Theta }
\newcommand{\nl }{\textsf{nleft}}
\newcommand{\nr }{\textsf{nright}}
\newcommand{\np }{\textsf{npow}}
\newcommand{\pd }{Pad\'{e} }

\newcommand{\PD }{Pad\'{e}}

\newcommand{\va }{\varphi }
\newcommand{\K }{M}
\newcommand{\R }{\text{Re}}
\newcommand{\Ip }{\text{Im}}
\newcommand{\q }{\quad }
\newcommand{\tr }{\theta_R }
\newcommand{\Om }{\Omega }
\newcommand{\nn }{{\nu{'}\gets \nu}}
\newcommand{\x }{\textbf }
\newcommand{\lm }{\lambda }

\begin{document}
\begin{frontmatter}

\title{Numerical Regge pole analysis of resonance structures in state-to-state reactive differential cross sections}

\author{ E. Akhmatskaya$^{a,c}$ and  D. Sokolovski$^{b,c}$}



\address[a]{Basque Center for Applied Mathematics (BCAM),\\ Alameda de Mazarredo 14,  48009 Bilbao, Bizkaia, Spain}

\address[b]{Department of Physical Chemistry, University of the Basque Country, Leioa, 48940, Spain}

\address[c]{IKERBASQUE, Basque Foundation for Science, Plaza Euskadi 5, 48009,  Bilbao, Bizkaia, Spain}


\begin{abstract}

This is the third (and the last) code in a collection of three programs [Sokolovski {\it et al} (2011), Akhmatskaya {\it et al} (2014)]  dedicated to the analysis of numerical data, obtained in an accurate simulation of an atom-diatom chemical reaction. 
Our purpose is to provide a  detailed description of a  \texttt{FORTRAN} code for complex angular momentum (CAM) analysis of the resonance effects in reactive angular scattering [for CAM analysis of integral reactive cross sections see [Akhmatskaya {\it et al} (2014)].
The code evaluates the contributions of a Regge trajectory (or trajectories) 
to a differential cross section in a specified range of energies. 
 The contribution is computed with the help of the methods described in   [Dobbyn {\it et al} (2007), Sokolovski and Msezane (2004),  Sokolovski {\it et al} (2007)]. Regge pole positions and residues are obtained by analytically continuing $S$-matrix element, calculated numerically for the physical integer values of the total angular momentum, into the complex angular momentum plane using the \texttt{PADE\_II}  program [Sokolovski  {\it et al} (2011)].
The code represents a reactive scattering amplitude as a sum of the components corresponding to a rapid "direct" exchange of the 
atom, and the various scenarios in which the reactants form long-lived intermediate complexes, able to complete several rotations before breaking up into products. The package has been successfully tested on 
the representative models, as well as the F + H$_2$$\to$ HF+H benchmark reaction. Several detailed examples are given in the text.

\begin{flushleft}

\end{flushleft}


\end{abstract}

\end{frontmatter}


{\bf PROGRAM SUMMARY}

\begin{small}
\noindent
{\em Manuscript Title:} Numerical Regge pole analysis of resonance structures in reactive state-to-state differential cross sections.               \\
{\em Authors:} E. Akhmatskaya, D. Sokolovski,                 \\
{\em Program Title:} \texttt{DCS\_Regge}                      \\
{\em Journal Reference:}                                      \\
{\em Catalogue identifier:}                                   \\
{\em Licensing provisions:}   Free software license                                \\
{\em Programming language:} \texttt{FORTRAN 90}                        \\
{\em Computer:} Any computer equipped with a \texttt{FORTRAN 90}
compiler\\
{\em Operating system:} UNIX, LINUX                           \\
{\em RAM:} 256 Mb                                       \\
{\em Has the code been vectorised or parallelised:} no                                      \\
{\em Number of processors used:} one                          \\
{\em Supplementary material:} \texttt{PADE\_II}, \texttt{MPFUN} and \texttt{QUADPACK}  packages, validation suites, script files, input files,
readme files, Installation and User Guide
{\em Classification:} Molecular Collisions                    \\
{\em External routines/libraries:} none \\ 
{\em CPC Program Library subprograms used:}   N/A          \\
{\em Nature of problem:}\\
 The package extracts the positions and residues of resonance poles 
 from numerical scattering data supplied by the user. This information is then used for the analysis of resonance structures observed in  reactive differential cross sections.
 \\
{\em Solution method:}\\
 The $S$-matrix element is analytically continued in the complex plane 
 of either energy or angular momentum with the help of \pd 
 approximation of type II. Resonance  Regge
trajectories are identified and their 
contributions to a differential cross section are evaluated at different angles and energies. 
   \\
{\em Restrictions:}\\
None.\\
  {\em Unusual features:}\\
  Use of multiple precision $MPFUN$ package.   \\
{\em Additional comments:} none\\
{\em Running time:}\\
from several minutes to hours depending on the number of energies involved, the  precision level chosen and the number of iterations performed.\\
{\em References:}
\begin{refnummer}
\item D. Sokolovski, E. Akhmatskaya and S. K. Sen, Comp. Phys. Comm. A, {\bf 182}  (2011) 448.
\item E. Akhmatskaya, D. Sokolovski, and C. Echeverr\'ia-Arrondo, Comp. Phys. Comm. A, {\bf 185}  (2014) 2127.
\item A.J. Dobbyn, P. McCabe, J.N.L. Connor and J.F. Castillo,
Phys. Chem. Chem. Phys., {\bf 1} (1999) 1115.
\item D. Sokolovski and A.Z. Msezane, 
Phys. Rev. A {\bf 70},  (2004) 032710.
\textcolor{black}{\item D. Sokolovski, D. De Fazio, S. Cavalli and V. Aquilanti,
Phys. Chem. Chem. Phys., {\bf 9} (2007) 5664.} 
\end{refnummer}

\end{small}

\newpage


\hspace{1pc}
{\bf LONG WRITE-UP}

\section{Introduction}

In the last fifteen years the progress in crossed beams experimental techniques has been matched by the  development  of state-of-the-art computer codes capable of modelling atom-diatom elastic, inelastic and reactive differential and  integral cross sections \cite{CODE1}- \cite{CODE2}. The differential cross sections (DCS), accessible to measurements in crossed beams, are often structured, and so offer a large amount of useful information about details of the collision or reaction mechanism. This information needs to be extracted and analysed, which often presents a challenging task.
One distinguishes two main types of collisions: in a {\it direct} collision the partners depart soon after the first encounter, while in a  {\it resonance} collision they form an intermediate complex 
(quasi-molecule), able to complete several rotations,  before breaking up into products. The resonance pathways may become important or even dominant at low collision energies. For this reason, accurate modelling and understanding of resonance effects gain importance in such fields as cold atom physics and chemistry of the early universe. 
\newline
Once the high quality scattered matrix is obtained numerically, one needs to understand the physics of the reaction, often not revealed until an additional analysis is carried out.
In particular, resonances invariably leave their signatures on the differential state-to-state cross sections, as the 
rotation of the intermediate complex can carry the collision partners into the angular regions not probed by the direct mechanism.
By a general rule of quantum mechanics, scenarios leading to the same outcome (in this case the same scattering angle) interfere,
and complex interference patterns can be produced in the reactive DCS. 
In this paper we propose and describe software for the analysis of such resonance patterns.
Relevant information on the Regge poles can be found in Refs.\cite{REGGE1}-\cite{REGGE2}. Some applications of the poles to the angular scattering and integral cross sections are discussed in Refs.\cite{DCS1}-\cite{DCS3}.
For a description of the type-II \PD\q approximation, used by the software, the reader is referred to Refs.\cite{PADE1},\cite{PADE2}. 
\section{Background and theory}
We start with a brief review of the concepts and techniques required for our analysis.
 \subsection{Reactive differential cross-sections}
For an atom-diatom reaction $A+BC\to AB+C$, a state-to-state differential cross section (DCS),  also called an angular distribution, 
 gives the number
of products, scattered at a given energy $E$  into a unit solid angle around a direction $\tr$, per unit time, per unit solid angle, for unit incoming flux of the reactants.  In the  entire-of-mass frame, $\tr$ is the angle between the initial velocity of the atom $A$ and that of a newly formed 
molecule $AB$. 
The  states of the molecule ($BC$ before and $AB$ after the reaction has taken place) is conveniently described 
 by the vibrational ($v$), rotational ($j$), and helicity ($\Om$ = projection of $j$ onto the final atom-diatom velocity) quantum numbers, 
 and the cross section is obtained as an absolute square of a {\it scattering amplitude} 
 \begin{eqnarray}\label{1}
 \sigma_{\nu^{\prime} \gets \nu}(\tr,E) = | f_{\nu^{\prime} \gets \nu}(\tr,E)|^2,\q \nu=(v,j,\Om).
 \end{eqnarray}
The scattering amplitude is given by a partial wave sum (PWS)
 \begin{eqnarray}\label{2}
 f_{\nu^{\prime} \gets \nu}(\tr,E) = ( ik_\nu)^{-1}\sum_{J=J_{min}}^\infty
 (J+1/2)
 S^J_\nn d^J_{\Om' \Om}(\pi -\tr),
 \end{eqnarray}
where $J$ is the total angular momentum, $S^J_\nn$ is a body-fixed scattering matrix element, 
$k_\nu$ is the initial translational wave vector of the reactants, and $d^J_{\Om' \Om}(\pi -\tr)$ stands for a reduced rotational matrix element
\cite{EDM}. The total angular momentum cannot be smaller than the largest of the two helicities, hence
$J_{min}=\text{max}(\Om,\Om')$. In the simpler case where both helicities are zero, $\Om=\Om'=0$, Eq.(\ref{2}) simplifies to
 \begin{eqnarray}\label{3}
 f_{\nu^{\prime} \gets \nu}(\tr,E) = ( ik_\nu)^{-1}\sum_{J=0}^\infty
 (J+1/2)
 S^J_\nn P_J(cos(\pi -\tr)),
 \end{eqnarray}
where $P_J(cos(\pi -\tr))$ is Legendre polynomial (see, e.g., \cite{BRINK}). 
In what follows we will restrict our analysis to the  special case (\ref{3}),  
although a similar approach can, in principle, be developed 
also for the transitions with non-zero helicity numbers \cite{GEN} ({\color {blue} see also \cite{NSR2})}.
Since $P_J(cos(\pi -\tr))=exp(i\pi J)P_J(cos(\tr))$, Eq.(\ref{3}) can be rewritten as
 \begin{eqnarray}\label{4}
 f_{\nu^{\prime} \gets \nu}(\tr,E) = ( ik_\nu)^{-1}\sum_{J=0}^\infty
 (J+1/2)
 \tilde S^J_\nn P_J(cos(\tr)),
 \end{eqnarray}
where $\tilde S^J_\nn\equiv \exp(i\pi J)S^J_\nn$.  
 With the $S$-matrix redefined in this manner, the PWS (\ref{4}) has the same form 
 as the one for the scattering amplitude in single-channel potential scattering.
 By the same token, a single channel amplitude (\ref{4}) can be rewritten in the form (\ref{3}), 
 so that a simple potential scattering model can (and will) be used to test our analysis. 
 \newline
In particular it is possible to crudely model chemical reactivity by evaluating first the scattering matrix for single-particle 
 potential scattering by a central potential $V(r)$, $S^J_{pot}$, and then constructing a "reactive $S$-matrix element" as
  \begin{eqnarray}\label{4b}
S^J=(-1)^J \exp(-J^2/\Delta J^2) S^J_{pot},\q |S^J|^2=\exp(-2J^2/\Delta J^2),
 \end{eqnarray}
where a Gaussian cut-off $\exp(-J^2/\Delta J^2)$  is introduced to mimic the decline in the probability  of exchanging the atom  $B$ as $J$ increases, and the reactants pass each other at ever greater distance. The DCS for two kinds of a single-channel "hard-sphere model"
($\delta(r)$ is a Dirac delta, 
and $\Omega >0$ {\color{black} is not to be confused with the helicity quantum number in Eq.(ref{1})} ),  

{\begin{eqnarray}\label{4a}
V(r) = \infty \q\q\q\q\q\q\q\q\q\q \text{for} \q r \le R-d,\q\q\q\q\q\q\q\q\n 
-V+ \Omega \delta (r-R) \q\q\q\q\text{for}\q R-d < r \le R,\q\q\q\q\q\q\n
 0\q\q\q\q\q\q\q\q\q\q\q\text{for} \q r> R,\q\q\q\q\q\q\q\q\q\q
 \end{eqnarray}
 
 where a hard core is surrounded by a narrow potential barrier at $r=R$, are shown in \autoref{plot:Fig1} a) and b). 
 The third panel c) in \autoref{plot:Fig1} shows the reactive DCS for the $F+H_2(0,0,0) \to HF(2,0,0) +H$ reaction, studied, e.g.,  in \cite{PCCP}. 


\begin{figure}[ht]
      \centering
      \subfloat{\includegraphics[angle=0,width=12.5cm]{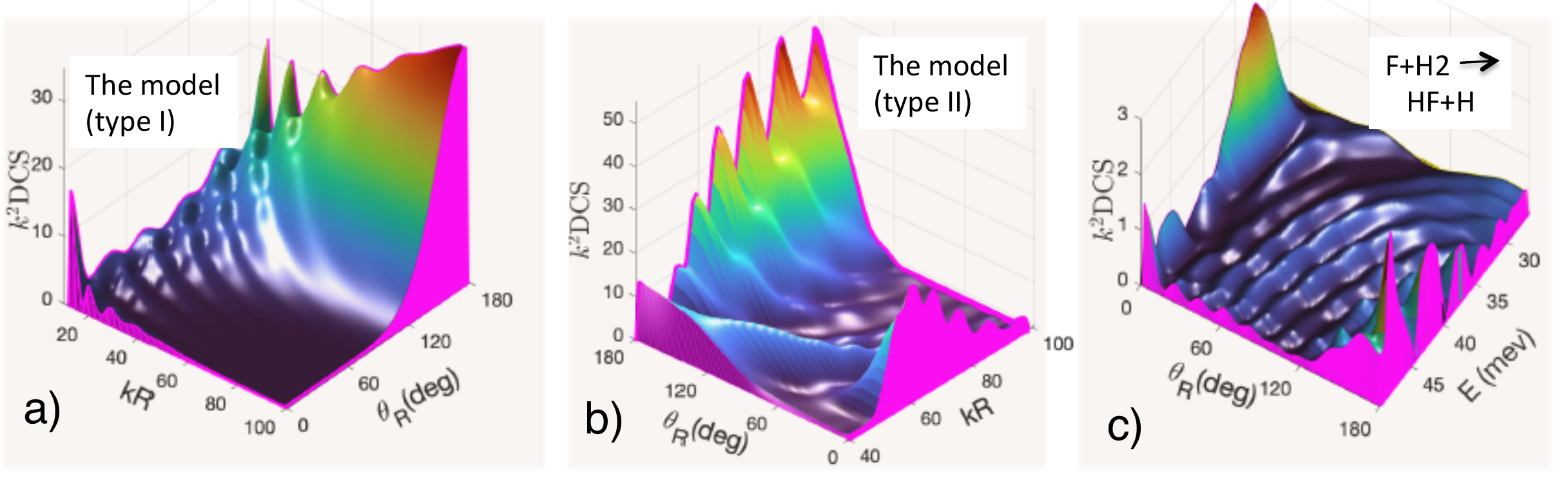}}
      \caption{ Dimensionless differential cross sections: a)  for the model (\ref{4a}) with $V<0$; 
b)  for the model (\ref{4a}) with $V=0$; and c) for the $F+H_2(0,0,0) \to HF(2,0,0) +H$ reaction \cite{PCCP}.
(Note the different orientations of the plots, chosen to reveal the features of interest, as discussed further in the text.)}\label{plot:Fig1}
\end{figure}
 The integral cross-sections for these three systems were analysed in \cite{ICS_REGGE}. Below we will use the same systems as examples 
 in our analysis of angular distributions. 
\subsection{Scattering "interferometry"}
The differential cross sections in \autoref{plot:Fig1} exhibit complicated interference patterns which may be used to gain further insight into what happens in the course of a reaction.
In fact, the DCS appear to be more sensitive to the details of the scattering mechanism that their  total (integral)  counterparts \cite{ICS_REGGE}. 
To see why, and in order to introduce useful terminology, we start by revisiting  scattering of a single {\it classical} particle by 
a short-ranged central potential $V(r)$ (see \autoref{plot:Fig2}a)). The trajectory of the particle with a given impact parameter lies in a plane, perpendicular to its angular momentum $J$, and the scattering angle $\theta$ is the angle between the particle's initial and final velocities $v_I$ and $v_F$.
Also relevant to our analysis, is the {\it winding angle} $\varphi$ swept by the vector $\vec R$, drawn to the particle from the origin, before it settles 
into its final direction $\vec R^F$. Clearly, the two are related by $\theta =\pi-\varphi$. 
Consider next a trajectory with a different impact parameter (not shown in \autoref{plot:Fig2}a)), which probes an attractive part of the potential, so that 
$\vec R$ rotates by $\va=\theta +\pi$. With a beam of particles incident on the potential, there will be a similar trajectory passing 
the scatterer on the other, "far" side, and ending in the same detector as the "nearside" (NS) trajectory with $\varphi=\pi-\theta$ (see \autoref{plot:Fig3}). 
Other possibilities include NS trajectories orbiting the potential $\K=0,1,2,...$ times, so that we have  $\varphi=\pi-\theta + 2\K\pi$, 
and the "farside" (FS) trajectories with $\varphi=\pi+\theta + 2\K\pi$. The presence, or otherwise, of such trajectories will, of course, depend on the properties 
of the potential, and the energy $E$ of the incident beam. 
\begin{figure}[ht]
\centering
\subfloat{\includegraphics[angle=0,width=12cm]{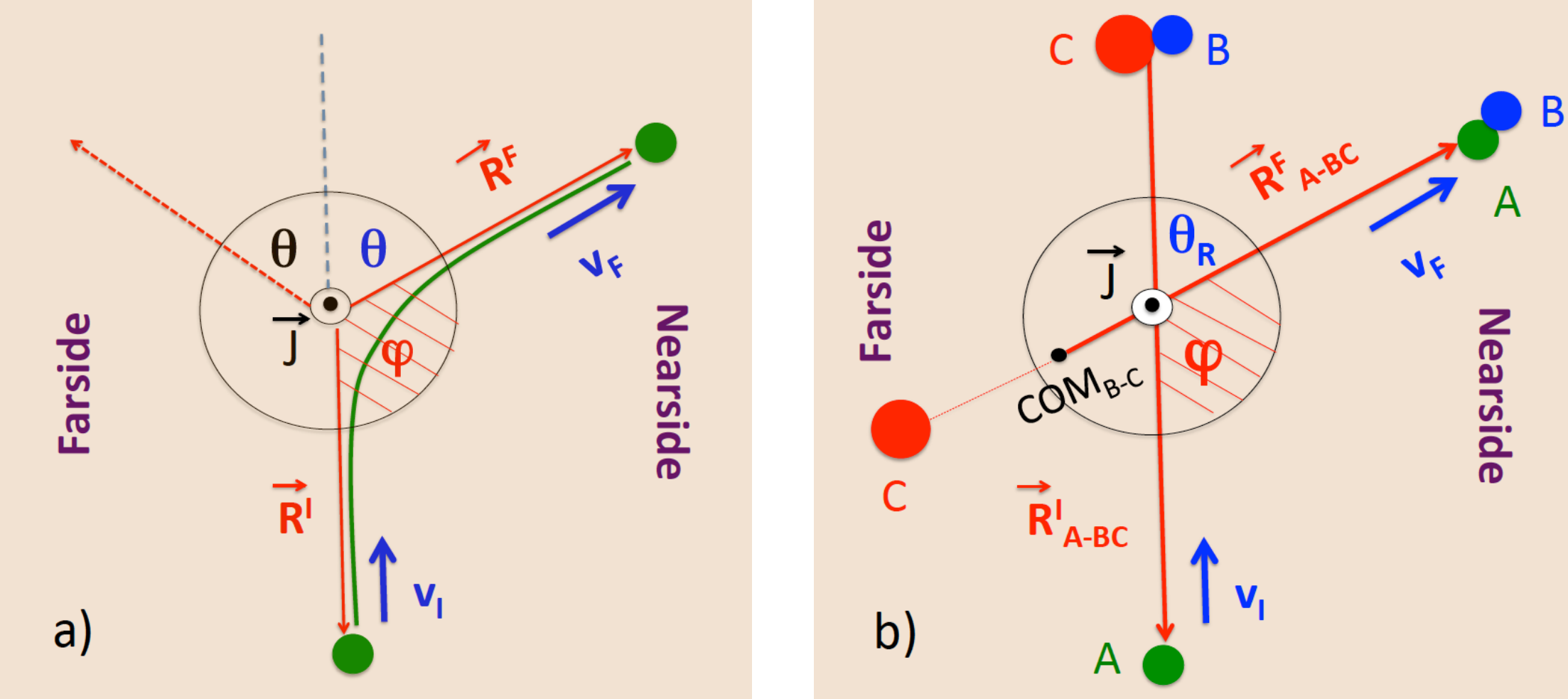}}
\caption{{} {\bf a)} A classical particle scattered by a central potential. In the course of the collision the particle's 
position vector $\vec R$ rotates by an angle $\va < \pi$, the scattering angle is $\theta= \pi-\va$, and the trajectory is of the $1$-st nearside type.
By symmetry, a similar trajectory (not shown, see \autoref{plot:Fig3}, for which   $\vec R$ (dashed) rotates by an angle $\va = \pi+\theta$, contributes to the 
same scattering angle, and is of the first far side type. In quantum mechanics, probability amplitudes for the two routes should be added, 
when calculating the probabilities.\label{plot:Fig2}
\newline
 { \bf b)} The initial and final arrangements for a reactive scattering ($A + BC \to AB + C$) into an angle $\theta_R$.
Projection of the reactants's Jacobi vector $\vec R_{A-BC}$  onto the plane, perpendicular to the total angular momentum 
$\vec J$, sweeps an angle $\va < \pi$, and the corresponding trajectory is of the $1$-st nearside type. }
\end{figure}
\begin{figure}[ht]
\begin{center}
\subfloat{\includegraphics[angle=0,width=6cm]{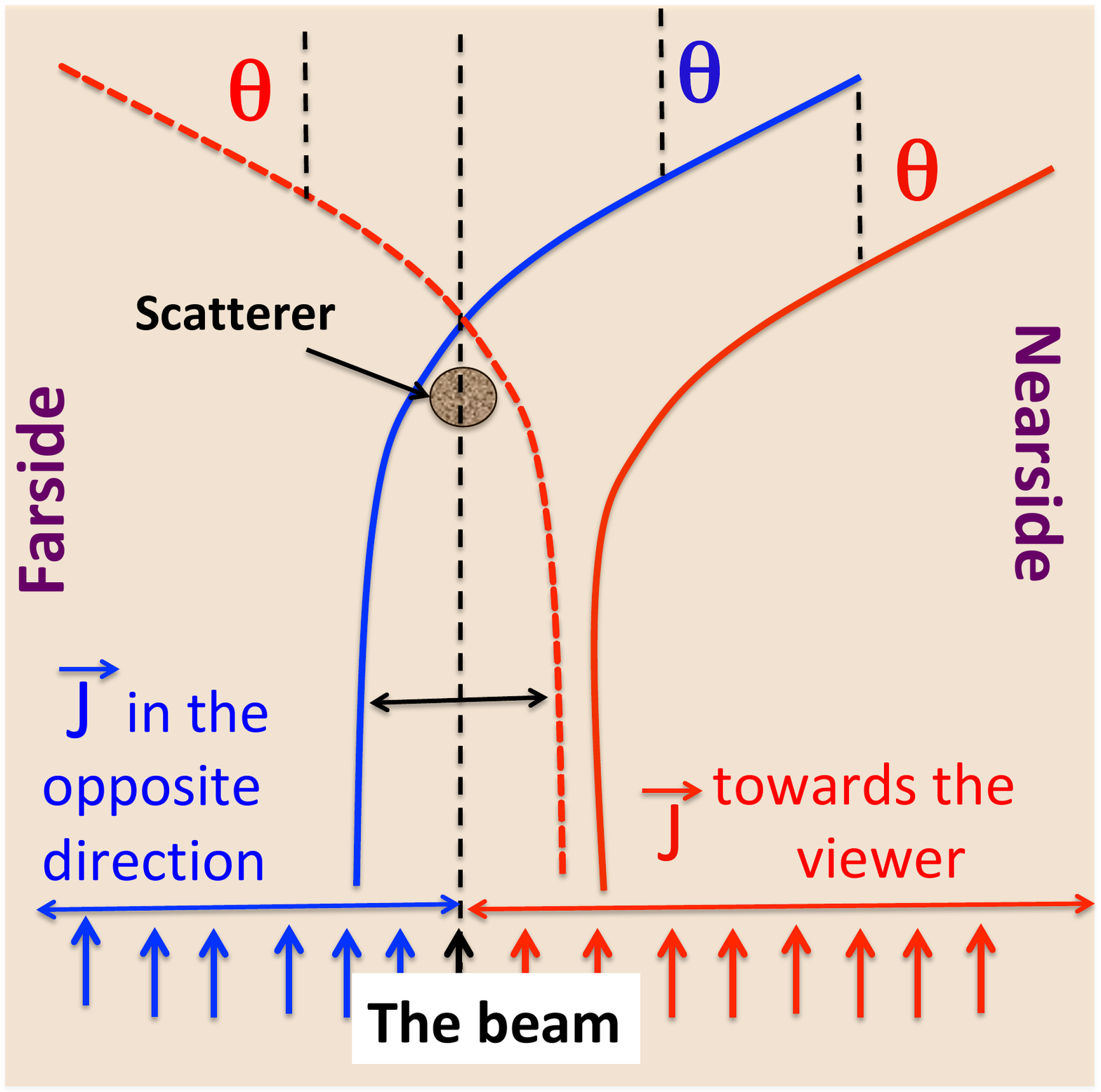}}
\end{center}
\caption{A beam of particle is incident on a central potential. Two trajectories (solid), passing on the opposite sides of the scatterer
with different impact parameters (angular momenta) 
experience repulsion and attraction, respectively, and are deflected into the same scattering angle $\theta$.
By symmetry, the one passing on the left of the target has a right hand side counterpart (dashed), corresponding 
to the same scattering angle. It is, therefore, sufficient (and convenient) to consider only the processes where a particle passes on the same side of the target, but may exit into the nearside, or the farside zone.  }\label{plot:Fig3}
\end{figure}
\newline
A classical analysis of atom-diatom reactive scattering turns out to be remarkably similar despite a larger number of variables involved 
 \cite{GEN}. For our purpose, it is sufficient to consider only the angle swept by the projection of the Jacobi vector $\vec R_{A-BC}$, drawn 
 from the COM of the $BC$ pair to the atom $A$, onto the plane perpendicular to the {\it total} angular momentum $\vec{J}$ (this now includes the orbital component $\vec{\ell}$, as well as the molecule's own angular momentum $\vec{j}$) \cite{GEN}. 
 In the zero-helicity case (\ref{3})-(\ref{4}), the initial and final directions of  $\vec R_{A-BC}$
 (although not necessarily its intermediate orientations)
are both perpendicular to $\vec{J}$, as shown in \autoref{plot:Fig2}b). Note that even though the atoms $B$ and $C$ are separated after the reaction, 
$\vec R_{A-BC}$ eventually settles into a final direction, opposite to the products' Jacobi vector $\vec R_{C-AB}$ drawn from the COM of $AB$ to $C$. 
As in the case of potential scattering, three-particle reactive trajectories can be classified according to the value of the winding angle $\va$, 
acquired in the course of the reaction. In particular, one has 
a {\it $\K+1$-st nearside reactive trajectory}, leading to a scattering angle $\theta_R$, if
 \begin{eqnarray}\label{5}
 \va^{NS}_\K= (2\K+1)\pi-\theta_R,\q 0< \theta_R< \pi,\q\K=0,1,2..., 
  \q\q\q
 \end{eqnarray}
 and a {\it $\K+1$-st farside reactive trajectory}, leading to a scattering angle $\theta_R$, if
\begin{eqnarray}\label{6}
 \va^{FS}_\K= (2\K+1)\pi+\theta_R,\q 0< \theta_R < \pi, \q \K=0,1,2...  \q\q\q
 \end{eqnarray}
\newline
Classically, the detection rate of particles (or diatomic molecules) scattered into an angle $\theta$ (or $\theta_R$)
is obtained by adding the numbers of particle travelling via all the trajectories  leading to the same angle $\tr$.
Quantally, one adds the {\it amplitudes} of all scenarios leading to the same outcome, and takes the absolute 
square of the result. The use of the interference patterns which may occur in a DCS,
in order to understand  the reaction's mechanism, is the main rationale behind the development 
of the software described in this article. 
\subsection{A simple nearside-farside decomposition of the scattering amplitude}
A simple method for separating the scattering amplitude (\ref{3})-(\ref{4}) into the nearside and farside components
originated in nuclear and heavy-ion physics \cite{BRINK}
was first applied to molecular collisions and chemical reactions  in 
\cite{NSR0}, \cite{NSR1} and \cite{NSR2}. It is based on approximating Legendre polynomials in the PWS (\ref{4}) by a sum of two 
components, 
\begin{eqnarray}\label{7}
P_J(cos(\tr))\approx P_J^+(\tr)+P_J^-(\tr), 
 \end{eqnarray}
where
\begin{eqnarray}\label{8}
P_J^\pm (\pi-\tr)\equiv [2\pi(J+1/2)\sin(\pi-\tr)]^{-1/2}\n
\times \exp\{\pm i[(J+1/2)(\pi-\tr)-\pi/4]\}.
 \end{eqnarray}
Accordingly, the scattering amplitude is written as a sum of two terms,
\begin{eqnarray}\label{9}
 f_{\nu^{\prime} \gets \nu}(\tr,E)=f^{NS}_{\nu^{\prime} \gets \nu}(\tr,E)+f^{FS}_{\nu^{\prime} \gets \nu}(\tr,E),
 \end{eqnarray}
where
\begin{eqnarray}\label{10}
 f^{NS,FS}_{\nu^{\prime} \gets \nu}(\tr,E)=( ik_\nu)^{-1}\sum_{J=0}^\infty
 (J+1/2)
 S^J_\nn P^{\pm}_J(\pi -\tr).
 \end{eqnarray}
 
 Since Eqs.  (\ref{7})-(\ref{8}) are valid in an angular range
 \begin{eqnarray}\label{11}
J\sin(\tr)>>1, 
 \end{eqnarray}
the approximation is useful in a "semiclassical limit",
where the PWS (\ref{9}) converges after sufficiently large number of terms
(in practice, $J_{max} \gtrsim 10$), and the reaction is dominated by large total angular momenta.
The possibility of extending the decomposition to low values of $J$ and non-zero helicities was discussed, for example, in \cite{NSR2},
but such an extension is beyond the scope of our analysis.)
The {\it simple} NS-FS decomposition (\ref{9}) inevitably fails for forward,  $\theta_R \approx 0$,  and backward, $\theta_R \approx \pi$ directions. 
\newline
In summary, an interference pattern appearing in $ f_{\nu^{\prime} \gets \nu}(\tr,E)$, with both $f^{NS}_{\nu^{\prime} \gets \nu}(\tr,E)$ and
$f^{FS}_{\nu^{\prime} \gets \nu}(\tr,E)$ remaining smooth function of $\tr$, 
indicates the importance of virtual scenarios in which 
the Jacobi vector $\vec R_{A-BC}$ rotates by an angle $\va> \pi$. 
\newline
However, the just described simple technique cannot distinguish between the scenarios in which $\vec R_{A-BC}$ completes 
several full rotations before settling into its final direction [cf. Eqs.(\ref{5}) and (\ref{6})], and can be improved further.
\subsection{A detailed nearside-farside decomposition of the scattering amplitude. Forward and backward scattering}
A further insight into the reaction's mechanism can be gained if the behaviour of the $S$-matrix element is known 
on the entire positive $J$-axis ($J\equiv |\vec J$). One can define a function $ S_\nn(\lambda)$, 
 \begin{eqnarray}\label{11a}
\lm\equiv J+1/2,
 \end{eqnarray}
analytic in the whole 
complex $\lm$-plane, such that the $ S^J_\nn$ in the PWS (\ref{3}) are its values 
at non-negative half-integer $\lm$'s, $ S^J_\nn= S_\nn(J+1/2)$, $J=0,1,2...$ 
(We will use either $J$ or $\lm$, whichever makes an expression look simpler). 
It can then be shown \cite{PCCP} that it is sufficient to know 
two functions of the winding angle $\va$, namely
\begin{eqnarray}\label{12}
\tilde f (\va)=\int_{0}^\infty \sqrt \lm  S_\nn(\lambda)\exp(i\lm \va)d\lm
 \end{eqnarray}
and 
\begin{eqnarray}\label{13}
\tilde g (\va)=\int_{0}^\infty \lm  S_\nn(\lambda)\exp(i\lm \va)d\lm,
 \end{eqnarray}
in order to be able to analyse the behaviour of the scattering amplitude in the whole angular range. 
\newline
In particular, with the help of the Poisson sum formula \cite{BRINK}, the NS and FS parts of the scattering amplitude (\ref{9}) can be written as \cite{PCCP}
\begin{eqnarray}\label{14}
 f^{NS}_{\nu^{\prime} \gets \nu}(\tr,E)=\sum_{\K=-\infty}^\infty  f^{NS}_{\nu^{\prime} \gets \nu}(\tr,E|\K)=\q\q\q\q\q\q\q\q\q\q\q\q\n
 ( ik_\nu)^{-1}[2\pi \sin(\tr)]^{-1/2}\sum_{\K=-\infty}^\infty \tilde f (\va^{NS}_\K)
 \exp[ -i(\K+1/4)\pi],\n
  f^{FS}_{\nu^{\prime} \gets \nu}(\tr,E)=\sum_{\K=-\infty}^\infty  f^{FS}_{\nu^{\prime} \gets \nu}(\tr,E|\K)=\q\q\q\q\q\q\q\q\q\q\q\q\n
  ( ik_\nu)^{-1}[2\pi \sin(\tr)]^{-1/2}\sum_{\K=-\infty }^\infty \tilde f (\va^{FS}_\K)
 \exp[ -i(\K+3/4)\pi],
 \end{eqnarray}
where the angles $\va^{NS}_\K$ and $\va^{FS}_\K$ are defined in Eqs.(\ref{5}) and (\ref{6}), respectively, 
and $\K$ is allowed to take negative values.
It is readily seen that the individual terms in the r.h.s. of Eqs.(\ref{14}) correspond to the processes in which 
$\vec R_{A-BC}$ undergoes $M$ complete rotations before settling into its final direction. The full NS or FS amplitude is, 
therefore, a result of interference between all such terms. Like the simple NS-FS decomposition (\ref{9}), 
Eqs.(\ref{14}) are best  used in the semiclassical case $J_{max} \gtrsim 10$, and for $\tr$ not too close to either $0$ or $\pi$.   
\newline
Similar decompositions of the amplitudes for small, $\tr \approx 0$, and large $\tr \approx \pi$ 
are obtained with the help of the second "unfolded amplitude", $\tilde g (\va)$, defined in Eq.(\ref{13}), 
\begin{eqnarray}\label{15}
 f_{\nu^{\prime} \gets \nu}(\tr\approx 0,E)=-\frac{[\tr/ \sin(\tr)]^{1/2}}{ \pi k_\nu}\sum_{\K=-\infty}^\infty(-1)^\K
 \times  
 \q\q\q \q\q\q\n 
 \int_{-\tr}^{\tr}\frac{\tilde g(\va+(2\K+1)\pi)}{\sqrt{\tr^2-\va^2}}d\va,\q\q\n
  f_{\nu^{\prime} \gets \nu}(\tr\approx \pi,E)=\frac{[(\pi-\tr)/ \sin(\tr)]^{1/2}}{ i\pi k_\nu}
  \sum_{\K=-\infty}^\infty(-1)^\K\times   
  \q\q\q \q\q\q\n
 \int_{-(\pi-\tr)}^{(\pi -\tr)}\frac{\tilde g(\va+2\pi\K))}{\sqrt{\tr^2-\va^2}}d\va.\q
 \end{eqnarray}
For the forward, $\tr= 0$, and backward, $\tr= \pi$, scattering amplitudes Eqs. (\ref{15}) yield \cite{PCCP}
\begin{eqnarray}\label{16}
 f_{\nu^{\prime} \gets \nu}(\tr=0,E)=\sum_{\K=-\infty}^\infty f^{FW}_{\nu^{\prime} \gets \nu}(E|M)=
 -k^{-1}_\nu\sum_{\K=-\infty}^\infty(-1)^\K 
\tilde g((2\K+1)\pi),
\end{eqnarray}
and
\begin{eqnarray}\label{17}
 {\color{black} f_{\nu^{\prime} \gets \nu}(\tr=\pi,E)=\sum_{\K=-\infty}^\infty f^{BW}_{\nu^{\prime} \gets \nu}(E|M)=}
   (ik_\nu)^{-1}
  \sum_{\K=-\infty}^\infty(-1)^\K
\tilde g(2\K\pi).
 \end{eqnarray}
Now the individual terms in Eqs.(\ref{16}) and (\ref{17}) correspond to the scenarios in which $\vec R_{A-BC}$
settles in the forward or backward direction after $M$ complete rotations. Note that both clockwise and anti-clockwise 
rotation are, in principle, possible. 
The regions of applicability of Eqs. (\ref{14}) and (\ref{15}) overlap \cite{FH2}, so that one has  representations of the relative scattering amplitude in the entire angular range $0\le \tr \le \pi$.
\newline 
The full dynamical information is, however, contained in the $S$-matrix elements and, therefore, in the amplitudes
$\tilde f (\va)$ and $\tilde g(\va)$, whose properties we discuss next. 
\subsection{Direct scattering amplitude. Primitive semiclassical approximation and the deflection function}
The forces acting between the collision partners are typically repulsive and, in the absence of other processes, 
one may expect that
 $\vec R_{A-BC}$ would rotate by an angle 
$\overline \varphi (J)$, $\pi \le \overline \varphi (J)\le 0$, 
where $ \overline \varphi (J)$ is uniquely defined by the value of $J$. 
 The small angular momenta correspond to a nearly head-on-collisions, 
where the collision partners bounce back after exchanging the atom $B$. 
In this case,  we have $\overline \varphi (J\approx 0) \approx 0$ and $\tr \approx \pi$. At  large angular momenta, $J >>1$, collision partners pass 
each other at a distance, and the transfer of the atom $B$ is suppressed. In this limit, we have $\overline \varphi (J >>1) \approx \pi$},
and $\tr \approx 0$. The $S$-matrix element can be written as $S_\nn(\lambda)=|S_\nn(\lambda)|\exp[i\Thet_\nn(\lm)]$, where
the first factor, $|S_\nn(\lambda)|_{\lm \to \infty}\to 0$,
for the reason just discussed. The rapidly changing phase $\Thet_\nn(\lm)$ determines the final orientation of $\vec R_{A-BC}$.
In particular, we have \cite{GEN}
\begin{eqnarray}\label{18}
d\Thet_\nn(\lm)/d\lm =-\overline \varphi (\lm). 
\end{eqnarray}
If $d\Thet_\nn(\lm)/d\lm$ happens to be a smooth decreasing function, relation (\ref{18}) can be inverted, 
to define angular momentum, $\overline \lm (\va)$, as a function of the unique winding angle $\va$. 
Evaluating the integrals (\ref{12}) and (\ref{13}) with the help of the  stationary phase approximation, we have
(for simplicity, we omit the hessians and various other factors, as these expressions are not used by our code)
\begin{eqnarray}\label{19}
\tilde f (\va)\sim \sqrt{\overline \lm (\va)}|S_\nn(\overline \lm (\va))|\exp[i\Thet_\nn(\overline \lm (\va))],\n
\tilde g (\va)\sim {\overline \lm (\va)}|S_\nn(\overline \lm (\va))|\exp[i\Thet_\nn(\overline \lm (\va))].\q
 \end{eqnarray}
From the previous discussion it is clear, that both functions will be contained mostly in the region $0 \le \va \le \pi$, 
with $|\tilde f (\va)|$ and $|\tilde g (\va)|$ decreasing as $\varphi$ increases.
We will call the region $0 \le \va \le \pi$ {\it the first nearside zone} and note, in addition,  that in this simple case 
the sums in Eqs.(\ref{14}) are reduced to just one term, 
\begin{eqnarray}\label{20}
 f_{\nu^{\prime} \gets \nu}(\tr,E)\approx ( ik_\nu)^{-1}[2\pi \sin(\tr)]^{-1/2} \tilde f (\va^{NS}_0)
 \exp[ -i\pi/4],
 \end{eqnarray}
and 
\begin{eqnarray}\label{21}
 f_{\nu^{\prime} \gets \nu}(\tr=0,E)\approx -k^{-1}_\nu 
\tilde g(\pi),\q\q\n
f_{\nu^{\prime} \gets \nu}(\tr=\pi,E)=(ik^{-1}_\nu) 
\tilde g(0)\approx 0.
 \end{eqnarray}
 One useful quantity is the {\it deflection function}, obtained by adding a $\pi$ to both sides of Eq.(\ref{18}), 
 which now yields the (unique) reactive scattering angle, for each value of the total angular momentum, 
\begin{eqnarray}\label{22}
\tr(\lm,E)= \pi -\frac{d\Thet_\nn(\lm)}{d\lm}.
 \end{eqnarray}
 A typical deflection function would, therefore, start from $\pi$ at $\lm\approx 1/2$, and then decrease to zero,
 often almost linearly, in the whole range of the angular momenta which contribute to the reaction \cite{G1}. 
 \newline
 The simple situation just described does not always occur in practice. 
 For example, in \autoref{plot:Fig1} direct scattering occurs only at higher energies ($kR \gtrsim 50$) in the panel a). 
 Deflection function exhibiting a minimum, or 
 crossing to negative value would indicate the presence of rainbow \cite{JNLCR1}-\cite{JNLCR3}, or glory \cite{JNLCG1}-\cite{JNLCG3}
 effects in the DCS. There is, however, an alternative universal language, relating such effects to the singularities of 
 the $S$-matrix element. We will consider them next.
 \subsection{Resonances and Regge poles}
\begin{figure}[ht]
\begin{center}
\subfloat{\includegraphics[angle=0,width=8cm]{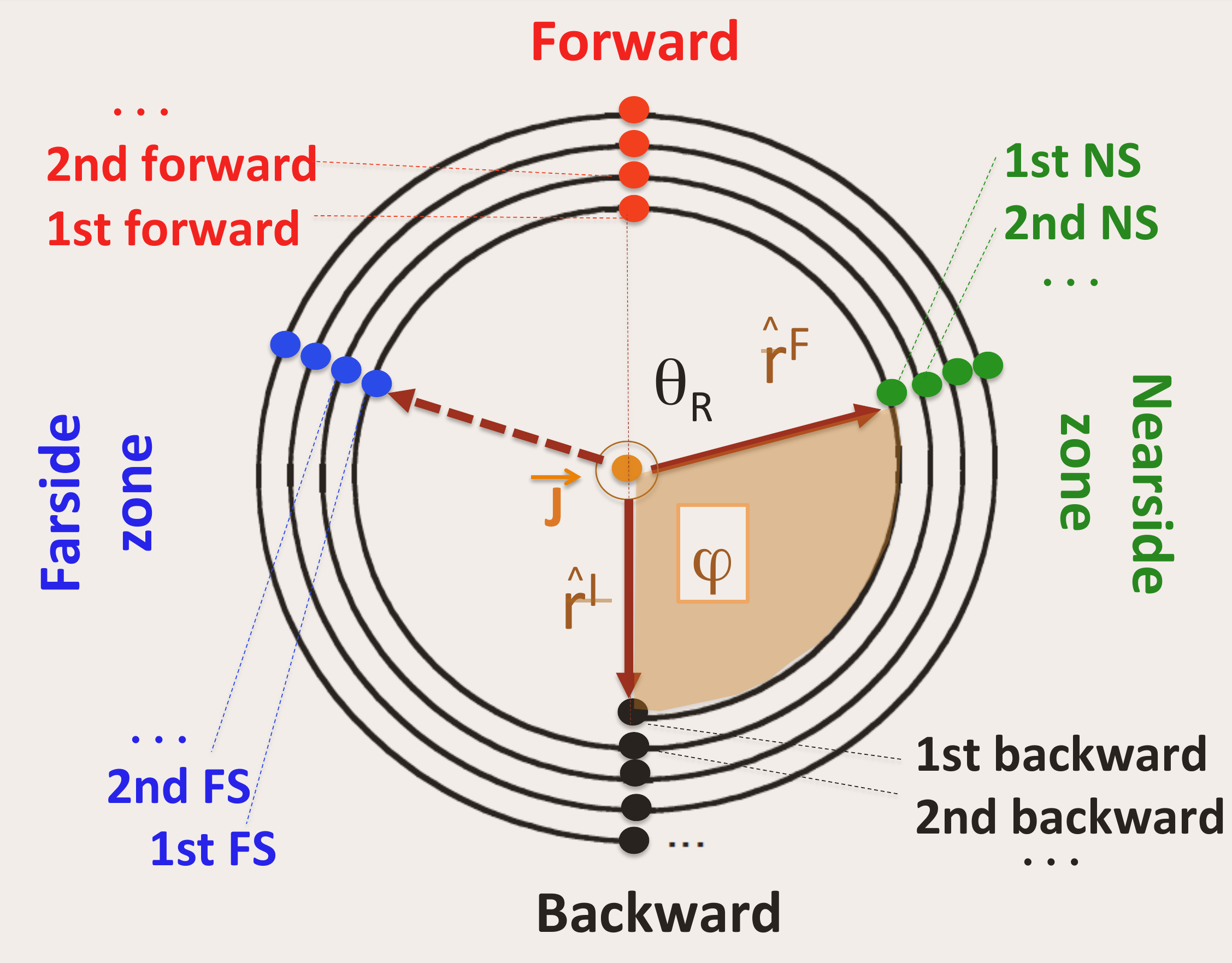}}
\end{center}
\caption{Possible rotations of a vector $\hat r \equiv \vec R_{A-BC}/|\vec R_{A-BC}|$ around the total angular momentum
$\vec{J}$, directed towards the viewer. In the semiclassical limit, the vector is expected to rotate in the positive sense, i.e., counter-clockwise.}\label{plot:Fig4}
\end{figure}

\begin{figure}[ht]
\begin{center}
\subfloat{\includegraphics[angle=0,width=12.5cm]{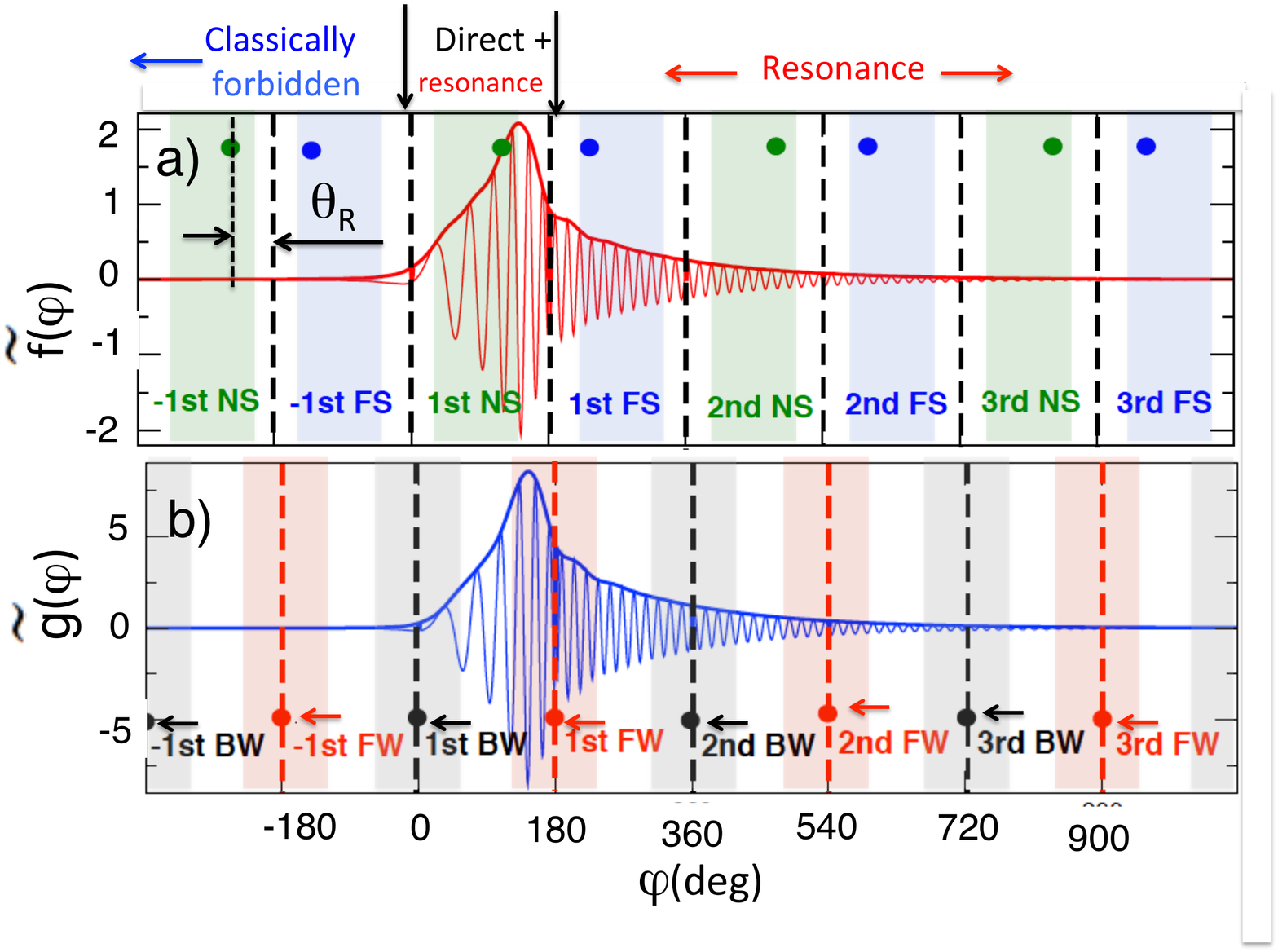}}
\end{center}
\caption{The "unfolded" scattering amplitudes:  a) the $\tilde f (\va)$ and b) $\tilde g (\va)$, for the
$F+HD(0,0,0)\to HF(3,0,0)+D$ reaction at $E=97.5$ $mev$ \cite{ICS_PCCP}. The PWS sum (\ref{3}) 
converges for $N_{max}\gtrsim 30$, and the semiclassical treatment applies. Also indicated are  (a)
various near- and farside zones, as well as (b) the winding angles contributing to the forward, ($\theta_R=0$) and
backward, ($\theta_R =\pi$), scattering amplitudes.}\label{plot:Fig5}
\end{figure}

\begin{figure}[ht]
\begin{center}
\subfloat{\includegraphics[angle=0,width=9cm]{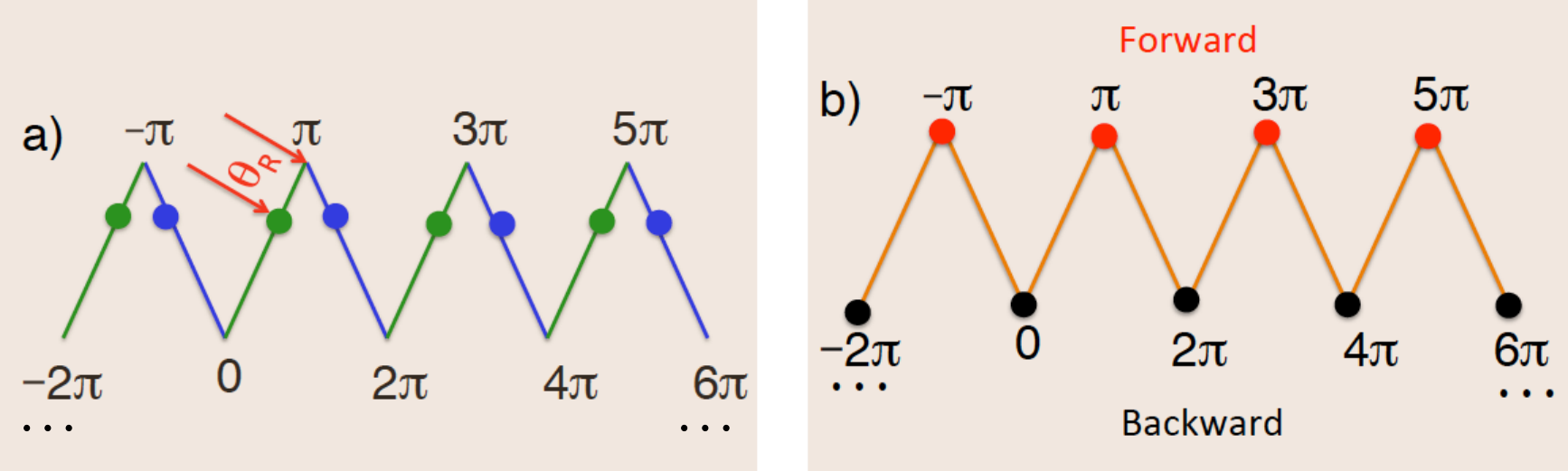}}
\end{center}
\caption{Folding back the unfolded amplitudes in \autoref{plot:Fig4}. a) The values of  $\tilde f (\va)$ for the $\va$'s corresponding to the 
same $\tr$ are added [with appropriate phase factors, cf. Eqs.(\ref{14})] to produce a sideway amplitude 
$f_\nn(\tr)$, $0 < \tr < \pi$. b) The values of $\tilde g (\va)$ for the $\va$'s corresponding to the 
same $\tr=0$ ($\tr=\pi$) are added [with appropriate signs, cf. Eqs.(\ref{16})-(\ref{17})] to produce a forward (backward) scattering amplitude 
$f_\nn(\tr=0)$ ( $f_\nn(\tr=\pi)$).}\label{plot:Fig6}
\end{figure}

 In our analysis, an important role is played by the (Regge) poles of the $S$-matrix,  $S_{\nu' \leftarrow \nu}(E,J)$ in the first quadrant of the complex $J(\lm)$-plane. A process in which the products form a long-lived triatomic "quasi-molecule", which breaks up into products after
 several complete rotations, manifests itself as a pole close to the real $J$-axis, in the first quadrant of the complex $J$-plane, 
 \begin{equation}\label{23}
S_{\nu' \leftarrow \nu}(E,J_n)=\infty, \quad Re J_n>0, \quad Im J_n>0.
\end{equation}
 The real part of the pole's position is related to the angular velocity, at which the quad-molecule spins. Its imaginary part, $Im J_n$,  defines the complex's 
 angular life - a typical angle by which it rotates before breaking up.
 \newline
  The complex must rotate in order to preserve the total angular momentum, 
 which causes the Jacobi vector $\vec R_{A-BC}$ to sweep a winding angle $\va > \pi$ as the complex decays into 
 the first farside, the second nearside, the second farside, and so on, zones. Typically (at least for $J_{max} >>j,j'$) the intermediate triatomic 
 needs to rotate in the positive sense around $\vec J$ (counter-clockwise, if $\vec J$ is pointing towards the observer), which explains why 
 the Regge poles are confined to the first quadrant of the complex $J$-plane. Indeed, it is easy to show \cite{PCCP} that, in presence of several ($N_{res}$)  resonance Regge poles at $J_n$ ($\lm_n=J_n+1/2$), the two functions in Eqs. (\ref{12}) and (\ref{13}), have a particularly simple form 
 beyond the first nearside zone, i.e., for $\va > \pi$,
\begin{eqnarray}\label{24}
\tilde f (\va) \approx \sum_{n=1}^{N_{res}}\tilde f^{(tail)}_n (\va)=
2\pi i \sum_{n=1}^{N_{res}}\sqrt{\lm_n}Res[S_{\nu' \gets \nu}(E,\lambda_n)]\exp(i\lm_n\va),\n
\tilde g (\va) \approx 
\sum_{n=1}^{N_{res}}\tilde g^{(tail)}_n (\va)=
2\pi i \sum_{n=1}^{N_{res}} {\lm_n}Res[S_{\nu' \gets \nu}(E,\lambda_n)]\exp(i\lm_n\va),\q
 \end{eqnarray}
 where $Res[S_{\nu' \gets \nu}(E,\lambda_n)]$ is the residue of the $S$-matrix element at the $n$-th Regge pole,
 \begin{eqnarray}\label{24a}
S_{\nu' \gets \nu}(E,\lambda_n) \to \frac {Res[S_{\nu' \gets \nu}(E,\lambda_n)]}{\lm-\lm_n}, \q \lm\to \lm_n=J_n+1/2.
 \end{eqnarray}
Thus, outside the first nearside zone, a pole creates an 
"exponential tail" $\sim \exp\{i\R[\lm_n]\va-\Ip[\lm_n]\va\}$, describing the rotation of a  triatomic complex,
accompanied by an exponential decay.
Note that a similar pole in the fourth quadrant of the complex $J$-plane, $\Ip[\lm_n]>0$, $\Ip[\lm_n] <0$,  would describe an unphysical increase, rather than a decay, and for this reason the physical poles have to lie above the real $J$-axis. In our semiclassical treatment, a capture into a long-lived intermediate state occurs at a total angle momentum $J \sim \R[J_n]$, corresponding to a direct scattering $\va_n$. In the vicinity of 
this winding angle direct scattering cannot be distinguished from the immediate decay of the newly formed complex \cite{FH1}.
At larger $\va$'s, still in the first NS zone, it may be possible to tell these contributions apart in  both $\tilde f (\va)$ and $\tilde g (\va)$. 
Although a reasonably simple description of  the behaviour of the two functions in the first NS zone can thus be obtained \cite{PCCP}, including it in a computer program is not straightforward, and
we will concentrate instead on the $\va \ge \pi$  angular range, where the direct component is usually absent [cf. Eqs.(\ref{22})]. 
Bearing in mind that the Jacobi vector $\vec R_{A-BC}$ rotates predominantly in the positive direction around the total angular momentum 
$\vec J$, we can start the summation  in Eqs.(\ref{14}),  (\ref{16}) and (\ref{17}) from $M=0$. 
With the help of Eq.(\ref{24})  the sideway (SW) scattering amplitude can be written as
\begin{eqnarray}\label{24a}
 f_{\nu^{\prime} \gets \nu}(0<\tr<\pi,E)\approx f^{NS}_\nn(\tr,E|0)+2\pi k_\nu^{-1}[2\pi \sin(\tr)]^{-1/2}\times
 \n
 \sum_{n=1}^{N_{res}}\sum_{K=0}^\infty
 \sqrt{\lm_n}Res[S_{\nu' \gets \nu}(E,\lambda_n)]\ 
 \exp[i(\lm_n\va_K-K\pi/2-\pi/4)]\n
 \equiv  f^{NS}_\nn(\tr,E|0)+\sum_{n=1}^{N_{res}}\sum_{K=0}^\infty f^{SW(tail)}_{\nu^{\prime} \gets \nu,n}(\tr,E|K)\q\q\n
  \equiv
  f^{NS}_\nn(\tr,E|0)+\sum_{n=1}^{N_{res}} f^{SW(tail)}_{\nu^{\prime} \gets \nu,n}(\tr,E),\q\q\q\q
\end{eqnarray}
where $\va_K=(-1)^{K+1} \tr +\pi\{K+[(-1)K+1]/2\}$, $K=1,2...$.
(Note that for an odd $K$, $\va$ lies in the $K$-th FS zone, whereas $f^{SW}_{\nu^{\prime} \gets \nu,n}(\tr,E|K)$ 
is what the $n$-th pole contributes in the $K$-th farside zone (see \autoref{plot:Fig4}).
Now the first term in Eq.(\ref{24a}) corresponds to a process in which $\vec R_{A-BC}$ rotates by a $\va < \pi$, 
while the second term includes the contributions from $N_{res}$ rotating complexes beyond the first NS zone, 
i.e., for $\va > \pi$.
\newline
By the same token, the forward (FW) and the backward (BW) scattering amplitudes, (\ref{16}) and (\ref{17}), can be approximated by 
\begin{eqnarray}\label{25}
 f_{\nu^{\prime} \gets \nu}(\tr=0,E)=
 -k^{-1}_\nu\sum_{n=1}^{N_{res}}\sum_{\K=0}^\infty(-1)^\K \tilde g^{(tail)}_n((2\K+1)\pi)\n
 \equiv \sum_{n=1}^{N_{res}}\sum_{\K=0}^\infty f^{FW(tail)}_{\nu^{\prime} \gets \nu,n,\K}(E)
 \equiv  \sum_{n=1}^{N_{res}}  f^{FW(tail)}_{\nu^{\prime} \gets \nu,n}(E)\q\q
\end{eqnarray}
and
\begin{eqnarray}\label{26}
  f_{\nu^{\prime} \gets \nu}(\tr=\pi,E)= (ik_\nu)^{-1}\left [\tilde g(0)
+\sum_{n=1}^{N_{res}}\sum_{\K=1}^\infty(-1)^\K
\tilde g^{(tail)}_n(2\K\pi)\right ]\q\q\n
\equiv f^{BW(direct)}_{\nu^{\prime} \gets \nu}
+\sum_{n=1}^{N_{res}}\sum_{\K=1}^\infty f^{BW(tail)}_{\nu^{\prime} \gets \nu,n,\K}(E)\equiv 
f^{BW(direct)}_{\nu^{\prime} \gets \nu}
+\sum_{n=1}^{N_{res}} f^{BW(tail)}_{\nu^{\prime} \gets \nu,n}(E).
 \end{eqnarray}
In Eq.(\ref{26}), $\tilde g(0)$ describes a direct recoil in a head-on collision, while $\tilde g((2\K+1)\pi)$, $\K \ge 0$, and $\tilde g(2\K\pi)$,  $\K \ge 1$, 
correspond to the processes in which a rotating complex decays into the $\tr=0$ and $\tr=\pi$ directions after several complete rotations.
A typical situation is shown in \autoref{plot:Fig5}, and \autoref{plot:Fig6} shows how the unfolded amplitudes $\tilde f(\va)$ and $\tilde g(\va)$
can be "folded back" to produce a scattering amplitude $ f_{\nu^{\prime} \gets \nu}(\tr,E)$ at a given angle $\tr$.
\newline
\newline

Geometric progressions in Eqs.(\ref{25}) and (\ref{26}) can be summed explicitly, to yield
\begin{eqnarray}\label{25a}
f^{FW(tail)}_{\nu^{\prime} \gets \nu,n}(E)= -\frac{ 2\pi i} {k_\nu} {\lm_n}Res[S_{\nu' \gets \nu}(E,\lambda_n)]\frac{\exp(i\pi \lm_n)}{1+\exp(2\pi i\lm_n)}
\end{eqnarray}
and
\begin{eqnarray}\label{26a}
f^{BW(tail)}_{\nu^{\prime} \gets \nu,n}(E)=-\frac{2\pi}{ k_\nu}{\lm_n}Res[S_{\nu' \gets \nu}(E,\lambda_n)]
\frac{1}{1+\exp(-2\pi i \lm_n)}.
 \end{eqnarray}
 Thus, Eq.(\ref{25a}) accounts for all forward contributions in \autoref{plot:Fig4}, while  Eq.(\ref{26a}) results from summing all 
 but the first backward contribution in \autoref{plot:Fig4}. 
\newline 
Similarly, summing the two geometric progressions (for the NS and FS contributions) yields


\begin{eqnarray}\label{27a}
f^{SW(tail)}_{\nu^{\prime} \gets \nu,n}(\tr,E)= \sqrt{\frac{2\pi}{ k_\nu^2\sin(\tr)}} {\lm_n^{1/2}}Res[S_{\nu' \gets \nu}(E,\lambda_n)]\times \n
\exp(-i\pi/4)\left [  \frac{\exp[i\lm_n(\pi-\tr)]} {1+\exp(-2\pi i\lm_n)}-i\frac{\exp[i\lm_n(\pi+\tr)]} {1+\exp(2\pi i\lm_n)} \right ].
 \end{eqnarray}
 Note that the amplitudes for the capture in a particular metastable state are always added,  and one cannot determine which complex has been formed just as one cannot know
which of the two slits was chosen by a particle in a Young's double-slit experiment \cite{FeynL}.
\newline
Finally, it is worth noting that in quantum mechanics, formation of a long-lived complex may involve classically forbidden processes, such as tunnelling across potential barriers. Such shape resonances, not seen in simulations employing classical trajectories,
will manifest themselves as Regge poles in the first quadrant of the CAM plane.
A description in terms of poles is also possible for rainbow \cite{G1} and glory \cite{HD2} effects, where 
a sequence of poles, rather just a single pole, are responsible for the structure in the deflection function (\ref{22}).
Application of the methods, described above in subsections 2.4-2.6 requires the knowledge of the behaviour of the $S$-matrix element 
in a subset of the complex $J$-plane, which is not readily provided by the scattering codes. Thus, an analytic continuation 
of $S_{\nu' \leftarrow \nu}(E,J_n)$ is a necessary element of our analysis. 
\subsection{ \pd reconstruction of the scattering matrix element}
The application of the equations (\ref{12})-(\ref{27a}) requires the knowledge of the pole positions $\lambda_n=J_n+1/2$, and the corresponding residues, $Res[S(E,\lambda_n)]$.  A computer code used in modelling  a chemical reaction typically evaluates the $S$-matrix elements, $S^J_\nn(E)$, for the physical integer values of $J=0,1,2,...N$, with $N$ sufficiently large to converge the partial waves sum (\ref{3}). Using these values, we construct a rational \pd approximant, ($[x]$ denotes integer part of $x$)
\newline
\begin{eqnarray}\label{27}
  S_{\nu' \gets \nu}^{Pade}(E,J)\equiv
 K_{N} \textrm{exp}[i(aJ^{2}+ bJ+c)]
 \times \frac{\prod_{i=1}^{[N/2]}(J-Z_i)}
 {\prod_{i=1}^{[(N-1)/2]}(J-P_{i})},
 \end{eqnarray} 
 where $P_i(E)$ and $Z_i(E)$ stand for poles and zeroes of the approximant, respectively, and $K_N(A)$, $a(E)$, $b(E)$ and $c(E)$ are energy dependent constants.
The approximant is conditioned to coincide with $S(E,J)$ at the $N$ integer values of $J$,
\begin{eqnarray}\label{27c}
 S_{\nu' \gets \nu}^{Pade}(E,J)=S^J_{\nu' \gets \nu}(E), \quad J=0,1,...N,
\end{eqnarray}%
and provides an analytic continuation for the exact function $S_{\nu' \gets \nu}(E,J)\approx  S_{\nu' \gets \nu}^{Pade}(E,J)$ in a region of the complex $J$-plane, containing the supplied values $S^J_{\nu' \gets \nu}(E)$. Inside this region, the poles and zeroes of the approximant serve
as good approximation to the true Regge poles and zeroes of the $S$-matrix element. 
The remaining poles and zeroes tend to mark the border of the region, beyond which \pd approximation fails \cite{G1},\cite{PADE2}.
Thus, for a given pole $J_n=P_j$, the residue is given by
\begin{eqnarray}\label{28}
Res[S_{\nu' \gets \nu}(E,\lambda_n)] =  K_{N} \textrm{exp}[i(aP_j^{2}+ bP_j+c)]
 \times \frac{\prod_{i=1}^{[N/2]}(P_j-Z_i)}
 {\prod_{i=1,i\ne j}^{[(N-1)/2]}(P_j-P_{i})},
 \end{eqnarray} 
which, together with Eq.(\ref{27})  provides all the data required for our analysis.
\newline
The program used to construct the \pd approximant {\color{black}(\ref{27})} is the  $\texttt{PADE\_II}$ code described  in the Refs. \cite{PADE2} and \cite{ICS_REGGE}. 
 \subsection{ A brief summary}
Our analysis consists in representing a reactive scattering amplitude   (\ref{3})  
by a sum  of simpler sub-amplitudes. It is used to relate the interference patterns observed in a reactive DCS  to the details 
of the reaction's mechanism, such as  capture of the collision partners into long- and not-so-long-lived intermediate states.  
The present approach can be applied in the case of zero initial and final helicities.  In particular, the  \texttt{DCS\_Regge}  code 
performs the following tasks. 
\newline
A) Evaluate the simple nearside-farside decomposition (\ref{9})-(\ref{10}) in a specified energy range.
\newline
B) Analytically continue the scattering matrix element into the complex $J(\lm$)-plane, using its values, $S^J_{\nu' \gets \nu}(E)$, 
at $J=0,1,...$. [The user should ensure that the input values correspond to $S^J_{\nu' \gets \nu}(E)$ in Eq.(\ref{3}), and not to 
$\tilde S^J_{\nu' \gets \nu}(E)= (-1)^J S^J_{\nu' \gets \nu}(E)$ in Eq.(\ref{4}).]
\newline
C) With the help of the values of $S_{\nu' \gets \nu}^{Pade}(E,J)$ on the real $J$-axis, evaluate the deflection function (\ref{22})
in a specified energy range.
\newline
D)  With the help of the values of $S_{\nu' \gets \nu}^{Pade}(E,J)$ on the real $J(\lm)$-axis, evaluate the "unfolded amplitudes"
$\tilde f (\va)$ and $\tilde g(\va)$ in Eqs. (\ref{12}) and  (\ref{13}), in a specified energy range.
\newline
E) Use Eqs.(\ref{24})  to relate the behaviour of the two amplitudes beyond the first nearside zone to the presence of Regge poles.
\newline
F) Implement the detailed nearside-farside decomposition (\ref{9}), (\ref{14}) of the sideways ($\tr$ is not too close to either $0$ or $\pi$)
scattering amplitude 
in a specified energy range. \
\newline
G) Implement the decompositions (\ref{25}), (\ref{26}) of the forward and backward scattering amplitudes, respectively.
{\color{black} \subsection {Limitations and some general remarks}
The method aims at giving  a clear picture of how the capture into metastable complexes, 
formed by the reactants, affects a DCS. It should, however, be noted that
\newline 
$\bullet $ The  \texttt{DCS\_Regge} code is designed for analysing {\it reactive} angular distributions, and uses the reactive scattering angle $\theta_R$ (cf.  \autoref{plot:Fig4}).
 \newline 
$\bullet $ The method applies to the systems for which the number of the partial waves included in the PWS is or order of $10$, or more.
Note that a large number of PWs may require the use of the multiple precision option, as explained below.
 \newline 
$\bullet$ The analysis is limited to the zero helicity case, where the initial and final directions of $\vec R_{A-BC}$ both lie in the plane, perpendicular to the total angular momentum $\vec J$ and, classically, $\theta_R$ can have any value between $0$ and $\pi$. This is no longer the case if one or both helicities have non-zero values \cite{GEN}. If so, $\theta_R$ is restricted to a narrower range, rotational matrices 
$d^J_{\Om' \Om}(\pi -\tr)$ develop caustics, and the semiclassical treatment becomes much more involved. Although a similar theory can, in principle, be developed, it is doubtful that its implementation be sufficiently straightforward. 
 \newline 
$\bullet$
In its present form, the method will not also distinguish between the direct and resonance scattering into the first  nearside zone, although the relevant theory can be found in \cite{PCCP}.
 \newline 
$\bullet$ Our analysis is only as good as the computation used to produce the $S$-matrix elements. It cannot compensate for any shortcoming of a close-coupling calculation used in its production. 
 \newline 
$\bullet$ However, if the $S$-matrix elements have been produced with a sufficient accuracy, so as to be in good agreement with the 
experimental data, our analysis would help to expose the details of the actual reaction mechanism, hidden in the PW decomposition of the scattering amplitude. For example, it would reveal a narrow resonance, falling between two integer values of the $J$, 
and explain the effect it has on the observed DCS.
 \newline 
$\bullet$ Over the years, similar methodology has been successfully applied to the $Cl+HCl$ \cite{HCL1,HCL2},  $F+H_2/HD$ \cite{FH1,FH2,FH3, PCCP} and $H+D_2$ \cite{HD1,HD2} benchmark systems. We expect the method to be well suited for a large number of various applications. 
 \newline 
$\bullet$ The analysis cannot reveal the physical origin of the resonance (e.g., its location in the entrance or exit channel on the potential surface, etc.), which must be established independently. }

\section {DCS\_Regge package: Overview} 
\subsection{ Installation}
This version of \texttt{DCS\_Regge} is intended for IA32 / IA64 systems running the Linux operating system. It requires \texttt{Fortran} and \texttt{C} compilers. 
The software is distributed in the form of a gzipped tar file, which contains the  \texttt{DCS\_Regge} source code, \texttt{PADE\_II 1.1} source code, \texttt{QUADPACK} source code, test suites for each package, as well as scripts for running and testing the code. The detailed structures of each subpackage, \texttt{DCS\_Regge}, \texttt{PADE\_II 1.1} and \texttt{QUADPACK}, are presented in the Appendices C, D and E.  
For users benefits we supply a file \texttt{README} for each package in directories  \texttt{DCS}, \texttt{DCS/PADE} and \texttt{DCS/QUADPACK}. The files provide a brief summary on the code structure and basic instructions for users. The  \texttt{DCS\_Regge} Manual (\texttt{DCSManual.pdf}), is located in the \texttt{DCS/} directory whereas \texttt{PADE\_II} Manual (\texttt{FManual.pdf}), can be found in the \texttt{DCS/PADE} directory.
In addition, the documents describing \texttt{PADE\_II} and \texttt{QUADPACK} are available from Mendeley Data (DOI: 10.17632/pt4ynbf5dx.1) and 
\newline
\texttt{http://www.netlib.org/quadpack/} respectively. 
Once the package is unpacked the installation should be done in the following order:

1.	Installing \texttt{QAUDPACK} library

2.	Installation of the \texttt{PADE\_II} package

3.	Installation of the \texttt{DCS\_Regge} package.  

Installation procedure for each sub-package is straightforward and can be successfully performed by following the instructions in 
\textcolor{black}{\texttt{DCS\_Regge} Manual. The procedure assumes using either Shell scripts (\texttt{QAUDPACK}) or a \texttt{Makefile\_UNIX} file (\texttt{PADE\_II}, \texttt{DCS\_Regge}) with the adjusted environmental variables.} 

\subsection {Testing}
Three test suits are prepared for each sub-package to validate the installation procedure. 

\subsubsection {Running the \texttt{QUADPACK} test}
To test the \texttt{QUADPACK} library, one has to run \texttt{quadpack\_prb.sh} script in the \texttt{/QUADPACK/scripts} directory. 

The results of 15 tests can be found in \texttt{quadpack\_prb\_output.txt} file in the 
\texttt{DCS/QUADPACK/test} directory. The message 
\newline
           \texttt{QUADPACK\_PRB: Normal end of execution}.
\newline
confirms that the code passed the validation test. 
\subsubsection {Running the \texttt{PADE\_II} test suite}
The input for 4 jobs,  \texttt{test1},  \texttt{test2},  \texttt{test3} and  \texttt{test4}, are provided in directories \newline
 \texttt{DCS/PADE/test/input/test\_name}, where \texttt{test\_name} is either  \texttt{test1} or  \texttt{test2} or  \texttt{test3} or  \texttt{test4}.
To submit and run a test suite, 
one should go to the directory \texttt{DCS/PADE/test} and run a script  \texttt{./run\_TEST}.
All tests will be run in separate directories. Each test takes about 1 - 5 minutes to run on a reasonably modern computer.
The message
\newline
\texttt{Test test\_name was successful}
\newline
appearing on the screen at the conclusion of the testing process confirms that the code passed the validation test \texttt{test\_name} . The results of the simulation can be viewed in the  \texttt{DCS/PADE/test/output/test\_name} directory.  
The message
\newline
 \texttt{Your output differs from the baseline!}
\newline
means that the calculated data significantly differ from that in the baselines. The files \texttt{PADE/test/output/test\_name/diff\_file} can be checked  to judge the differences. 
\subsubsection {Running the \texttt{DCS\_Regge} test suite}
For simplicity, the test suite for  \texttt{DCS\_Regge} is designed in the similar manner as the test suite for  \texttt{PADE\_II}. 
The input for 3 jobs,  \texttt{test1},  \texttt{test2} and  \texttt{test3} are provided in directories \texttt{DCS/test/input/test\_name} respectively, where  \texttt{test\_name} is either  \texttt{test1} or  \texttt{test2} or  \texttt{test3}. 
The test suit can be run in the directory \texttt{DCS/test} using the following commands:
 \texttt{./run\_TEST}.
All tests are run in separate directories, \texttt{DCS/test/output/test1}, \texttt{DCS/test/output/test2} and \texttt{DCS/test/output/test3}.
Each test takes about 1 minute to run on a reasonably modern computer.
To analyse the test results, the message on the screen at the completion of the testing process has to be inspected. 
The message
\texttt{ Test test\_name was successful}
confirms that the code passed the validation test  \texttt{test\_name}. The results of the simulation can be viewed in the \texttt{DCS/test/output/test\_name} directory.  
The message

        \texttt{test\_name  output differs from the baseline!}
        
\textcolor{black} {
        \texttt{Check your output in output/test\_name/basel}}

\textcolor{black} {
        \texttt{Baseline file: baselines/test\_name/basel}}

        \texttt{Differences: output/test\_name/error\_file}

means that the calculated data significantly differ from that in the baselines. The differences can be found in the files  \texttt{DCS/test/output/test\_name/error\_file}.
%
\section {Computational modules of DCS\_Regge} 
\subsection {\pd reconstruction and $\texttt{PADE\_II}$ options}
As discussed above, an important part of the calculation consists in performing an analytical continuation of the $S$-matrix element into the CAM plane.
The user has some flexibility in doing so. It concerns mostly the quadratic phase in Eq. (\ref{27}), which must itself be determined in the course of the \pd reconstruction. The need for separating this rapidly oscillating term arises from the fact that the \pd technique used here, works best for slowly varying functions. Thus, by removing the oscillatory component, one expands  the region of validity of the approximant in the CAM plane, which allows for the correct description of a larger number of poles.
The extraction of the quadratic phase proceeds iteratively. Since sharp structures in the phase of the $S$-matrix element usually come from the Regge poles and zeroes located close to the real axis, one defines a strip $\texttt{-dxl}<Im J<\texttt{dxl}$ around the real $J$-axis, and removes from the previously constructed approximant all poles and zeroes inside the strip. The smoother phase of the remainder is fitted to a quadratic polynomial, and this quadratic phase is then subtracted from the phase of the input values of $S$, after which a new approximant is constructed with these modified input data. The process is repeated \texttt{niter} times resulting in a (hopefully) improved \pd approximant.
There is no rigorous estimate of the improvement achieved, and the practice shows that in many cases using \texttt{niter  > 1} gives tangible benefits, while in some cases better results are achieved with \texttt{niter}=$1$ or $2$. It is for the user to decide on the best values of \texttt{dxl} and \texttt{niter} for a particular problem.
\newline
The input files required for running the \texttt{\PD\_II} code are stored in the input directory where they are labelled $1,2...N_E$.
A typical input file is given in  \autoref{plot:Fig12} (Appendix B). The file differs from the similar input used in  \texttt{\PD\_II} package reported in \cite{PADE2} by one line added at the end, which should contain the collision energy in $meV$, and is identical to an  input file used by \texttt{ICS\_Regge} \cite{ICS_REGGE}. 
\newline
Other parameters for \pd reconstruction are read from the \texttt{input/INPUT} file. 
\newline
The entry \textcolor{black}{(\#16)} determines whether there should be a change of parity from the original data. This depends on the convention used in calculating the $S$-matrix elements, as explained in \cite{PADE2} (we use $0$ for \texttt{no} and $1$ for \texttt{yes}). The recommended 
value is $1$.
\newline
Entry \textcolor{black}{(\#17)} decides whether 
one should remove the guessed values of the quadratic phase prior to the construction of the first \pd approximant in a series of iterations.  Its recommended value is $1$ (\texttt{yes}). 
\newline
Entry \textcolor{black}{(\#18)} determines if multiple precision routines should be used in calculating the \pd approximant. The recommended value is $1$ if the number of partial waves (PW) exceeds $40$.
It can also be used for a smaller number of (PW) to check the stability of calculations.
\newline
Entries \textcolor{black}{(\#19 and  \#21)} allow us to repeat the calculations with added non-analytical noise of magnitude \texttt{fac}, \texttt{nstime} times. This may be needed to check the sensibility of calculations to numerical noise. The recommended initial values are \texttt{nstime=0} and \texttt{fac=0.00000001}, in which case no noise is added.
\newline
Entry \textcolor{black}{(\#20)} determines the number of points in the graphical output from \texttt{\PD\_II} \cite{PADE2}.
\newline
Finally, the user has the options of changing  the number of partial waves, the number of iterations and the value of \texttt{dxl} for all files used in the current run by setting to $1$ \texttt{iover1}, \texttt{iover2} and \texttt{iover3} in entries \textcolor{black}{(\#22, \#23 and  \#24)}. The corresponding parameters are reset to the values \texttt{nread1}, \texttt{niter1}, and \texttt{dxl1}, specified in the entries \textcolor{black}{(\#25, \#26 and \#27)}, respectively.

\subsubsection{Changes made to $\texttt{PADE\_II}$ }
The changes from the previous version \cite{PADE2} include 
\newline
(I) replacement of all the Numerical Algorithms Group (NAG) routines with ones available in the public domain, and 
\newline
(II) 
provision of additional controls allowing to change the parameters of \pd reconstruction for all energies in the run at once, without changing individual input files labelled $1$,$2$,..., as discussed in the previous Section. 
\newline
A brief summary of the changes made to subroutines is given below.
\newline
{\it Subroutine} \texttt{FIT} {\it(\texttt{fit.f})}
\newline
The NAG routine \texttt{g05ccf} has been replaced by a subroutine \texttt{svdfit} described in section 15x.4 of {\it Numerical Recipes in C: The Art of Scientific Computing (Second Edition), published by Cambridge.}
\newline
\newline
{\it Subroutine} \texttt{ZSRND} {\it(\texttt{zsrnd.f})}
\newline
The NAG routine \texttt{E02ACF}  has been replaced by a sequence of calls to the system routines \texttt{srand48}  and \texttt{drand48}.
\newline
\newline
{\it Subroutine} \texttt{IGET\_SEED} {\it(\texttt{iget\_seed.c})}
\newline
Added new routine generating the seed for \texttt{srand48}.
\newline
\newline
{\it Wrapper (\texttt{wrapper.c})}
\newline
Added wrapper allowing for calling C-routines \texttt{IGET\_SEED}, \texttt{srand48} and \texttt{drand48} in a \texttt{Fortran} code.
\newline
\newline
{\it Subroutine} \texttt{FINDPOLES\_NAG} {\it(\texttt{findpoles\_NAG.f})}
\newline
Removed.
\newline
\newline
{\it Subroutine} \texttt{FINDZEROS\_NAG} {\it(\texttt{findzeros\_NAG.f})}
\newline
Removed.
{\subsection{The structure of $ \texttt{DCS\_Regge}$ } 

The $ \texttt{DCS\_Regge}$ application is a sequence of 24 FORTRAN files. The files are listed below, and their functions are explained.
\newline
\newline
{\it Program \texttt{DCS\_Regge} (\texttt{DCS\_Regge.f})}
\newline
Main program.
\newline
\newline
{\it Subroutine \texttt{OPEN\_IO} (\texttt{open\_io.f})}
\newline
Opens the files.
\newline
\newline
{\it Subroutine \texttt{CLOSE\_IO} (\texttt{close\_io.f})}
\newline
Closes the files.
\newline
\newline
{\it Subroutine \texttt{READ1} (\texttt{read1.f})}
\newline
Reads the original input file at a given energy and the file \texttt{screen.pade}.
\newline
\newline
{\it Subroutine \texttt{READ} (\texttt{read.f})}
\newline
Reads the parameters of the \pd  reconstruction at a given energy.
\newline
\newline
{\it Subroutine \texttt{SORT} (\texttt{sort.f})}
\newline
At a given energy, selects poles and zeroes in the specified region of the complex angular momentum ($J$-) plane and discards pole/zero pairs (Froissart doublets).
\newline
\newline
{\it Subroutine $\texttt{DCS\_exnearfar}$ ($\texttt{dcs\_exnearfar.f}$)}
\newline
Calculates the exact DCS (\ref{3}) and its simple nearside-farside decomposition (\ref{9}), for a specified range of energies.
\newline
\newline
{\it Subroutine \texttt{PHASE} (\texttt{phase.f})}
\newline
Calculates the deflection function, and the phase \textcolor{black}{$\Thet_\nn (J,E)$} of the $S$-matrix element  $S_\nn(\lm,E)
=|S_\nn(\lm,E)|\exp[i\Theta_\nn (J,E)]$,
for a specified range of energies. 
\newline
\newline
{\it Subroutine \texttt{DCS\_fg} (\texttt{dcs\_fg)}}
\newline
Evaluates the unfolded amplitudes $\tilde f (\va)$ and $\tilde g (\va)$ for specified ranges of the winding angle 
$\va$ and energy E.
\newline
\newline
{\it Subroutine \texttt{DCS\_side} (\texttt{dcs\_side.f})}
\newline
Evaluates the detailed near- and farside amplitudes  $f^{NS}_{\nu^{\prime} \gets \nu}(\tr,E|\K)$ and 
$f^{FS}_{\nu^{\prime} \gets \nu}(\tr,E|\K)$ in Eqs.(\ref{14}), for a given reactive scattering angle 
$\tr$, and specified ranges of $\K$ and energy $E$.
\newline
\newline
{\it Subroutine \texttt{DCS\_forb} (\texttt{dcs\_forb.f})}
\newline
Evaluates the decompositions of the forward and backward scattering amplitudes into 
$f_{\nu^{\prime} \gets \nu}(\tr=0,E|M)$ and $f_{\nu^{\prime} \gets \nu}(\tr=\pi,E|M)$ 
[cf. Eqs.(\ref{16}) and (\ref{17})] for specified ranges of $\K$ and energy $E$.
\newline
\newline
{\it Subroutine \texttt{REGGE1} (\texttt{regge1.f})}
\newline
Evaluates, for a single (first) energy $E$,  the exponential tails in Eq.(\ref{24}), $\tilde f^{tail} (\va|n)$ and $\tilde g^{tail} (\va|n)$, produced 
by a particular Regge pole at $\lm=\lm_n$ (chosen by the user) in an angular range beyond the 1-st NS zone,  $\va \ge \pi$. 
\newline
\newline
{\it Subroutine \texttt{REGGE2} (\texttt{regge2.f})}}
\newline
\textcolor{black}{For a specified range of energies E, follows the Regge trajectory, to which the pole at $\lm=\lm_n$, 
previously chosen by the user during the execution of \texttt{REGGE1}, belongs.} Full contributions of the pole 
to the forward, backward, and sideway scattering amplitudes, (\ref{25a}), (\ref{26a}), and (\ref{27a}), are evaluated
at each $E$, as well as the individual terms of the geometric progressions.  
\newline
\newline
{\it Function \texttt{ZPADE} (\texttt{zpade.f})}
\newline
\textcolor{black}{Calculates the full \pd approximant.}
\newline
\newline
{\it Function \texttt{ZPADE1} (\texttt{zpade1.f})}
\newline
\textcolor{black} {Calculates the full \pd approximant \textcolor{black}{in a different manner.}}
\newline
\newline
{\it Function \texttt{ZPR} (\texttt{zpr.f})}
\newline
Evaluates the ratio of the two polynomials in the \pd approximant.
\newline
\newline
{\it Function \texttt{ZRES} (\texttt{zres.f})}
\newline
Calculates (part of)  the residue for a chosen Regge pole from the \pd approximant.
\newline
\newline
{\it Function \texttt{ZEX} (\texttt{zex.f})}
\newline
Evaluates the partial wave sum (\ref{3}).
\newline
\newline
{\it Function \texttt{ZNE} (\texttt{zne.f})}
\newline
Evaluates the nearside partial wave sum, $f^{NS}_\nn(\tr,E)$ in Eq.(\ref{10}).
\newline
\newline
{\it Function \texttt{ZFA} (\texttt{zfa.f})}
\newline
Evaluates the farside partial wave sum, $f^{FS}_\nn(\tr,E)$ in Eq.(\ref{10}).
\newline
\newline
{\it Subroutine \texttt{LPFN} (\texttt{lpfn.f})}
\newline
Evaluates Legendre polynomials $P_J (\pi-\tr)$  in Eq.(\ref{3}).
\newline
\newline
{\it Subroutine \texttt{FNFN} (\texttt{fnfn.f})}
\newline
Evaluates $P_J^\pm (\pi-\tr)$ in Eq.(\ref{8}).
\newline
\newline
{\it Subroutine \texttt{FOURIER} (\texttt{fourier.f})}
\newline
Evaluates the Fourier transforms (\ref{12}) and (\ref{13}).
\newline
\newline
{\it Function \texttt{FST3} (\texttt{fst3.f})}
\newline
Supplies the integrand for the integral evaluated in \texttt{FOURIER}.
\newline
\newline
There are two additional utilities.
\newline
\newline
{\it Subroutine \texttt{SKIP} (\texttt{skip.f})}
\newline
Decides which of the input data/energies must be included in the current run.
\newline
\newline
{\it Subroutine \texttt{SUBTR} (\texttt{subtr.f})}
\newline
Subtracts from a forward, sideway, or backward scattering amplitude the resonance contribution from the  Regge pole trajectory
used by \texttt{REGGE2}.

All files are located in the \texttt{DCS/src} directory.
%


\subsection{The  $\texttt{QUADPACK}$ library }
$\texttt{QUADPACK}$ is a $\texttt{FORTRAN}$ subroutine package for the numerical computation of definite one-dimensional integrals. It originated from a joint project of R. Piessens and E. de Doncker (Appl. Math. and Progr. Div.- K.U.Leuven, Belgium), C. Ueberhuber (Inst. Fuer Math.- Techn.U.Wien, Austria), and D. Kahaner (Nation. Bur. of Standards- Washington D.C., U.S.A.) \cite{E1}.
\newline (http://www.netlib.org/quadpack/).

Currently one library subroutine, $\texttt{DQAGS}$, is used in the $\texttt{DCS\_Regge}$. The subroutine estimates integrals over finite intervals using an integrator based on globally adaptive interval subdivision in connection with extrapolation \cite{E2} by the Epsilon algorithm \cite{E3}. The subroutine is called from the $\texttt{DCS\_Regge}$ subroutine $\texttt{FOURIER}$. 
For users convenience the whole library is available in the package $\texttt{DCS\_Regge}$.  The link to the library is provided in 
$\texttt{Makefile\_UNIX}$ in $\texttt{DCS/src}$.

\section{Running the  \texttt{DCS\_Regge} code}
\subsection {Creating input data}

Two types of input files 
are required for running calculations: a parameter file, \texttt{INPUT}, located in the \texttt{DCS/input} directory, and a file or a set of files for running  \texttt{PADE\_II}. The latter
has(ve) to be supplied in \texttt{DCS/input/PADE\_data} directory and it (or they) contain(s) the data to be \pd approximated. The name of directory \texttt{PADE\_data} can be chosen arbitrary and should be specified in the parameter file \texttt{INPUT} before starting the calculations. The names of the input files in the directory \texttt{PADE\_data} are fixed to be $1,2,.. N_E$. Each file contains the input parameters required to run \texttt{PADE\_II}, previously computed values of $S_{\nu' \gets \nu}(E,J)$ for $J=0,1,2, ...J^{max}_i$, for the energy $E_k$, $k=1,2,..N_E$, and the value of the energy itself. 
An example of such an input file is given in Appendix B and also provided in the directory \texttt{DCS/input}. 
\newline
The file \texttt{DCS/input/INPUT} is self-explanatory and describes each input parameter to be 
specified. Please notice, that each input entry appears between colons (:). For an example of the parameter file \texttt{INPUT} see Appendix A. 

We provide input files for all test cases considered in this study under the names  
\newline
\texttt{DCS/input/INPUT.BOUND.DCS},  
\newline
\texttt{DCS/input/INPUT.BOUND.DCS.30}, 
\newline
\texttt{DCS/input/INPUT.META.DCS},
\newline
\texttt{DCS/input/INPUT.META.DCS.60},
\newline
\texttt{DCS/input/INPUT.FH2},
 and  
\texttt{DCS/input/INPUT.FH2.DCS.32.27}.
\newline 

\subsection {Executing \texttt{DCS\_Regge}}
The script \texttt{runDCS} in the \texttt{DCS/} directory automates calculations. The following assumptions are made in the script:
\newline
*  all binaries for \texttt{DCS\_Regge} are placed in \texttt{DCS/bin}, whereas the binaries for the \texttt{PADE\_II} package are located in \texttt{DCS/PADE/bin}.
\newline
*  input files are located in the directory \texttt{DCS/input}. The names of the input files are chosen as described in section 5.1.
\newline
*  output files can be found in \texttt{DCS/output} on completion of a calculation.

We recommend running a calculation in the directory \texttt{DCS/}. The command
\texttt{./runDCS}
immediately starts a calculation.
\subsection {Understanding the run script \texttt{runDCS}}
The run script \texttt{runDCS} located in the \texttt{DCS/} directory does not require any tuning, editing or corrections in order to start a calculation. Provided that the parameter file \texttt{DCS/input/INPUT} is prepared for calculations, the run script \texttt{runDCS} takes care
of the following steps in the following order:
\newline
1. INITIALIZATION
\newline
* Edits input parameter file \texttt{INPUT}
\newline
* Reads input parameters from \texttt{INPUT}
\newline
* Prepares directories for runs, i.e. sets useful directories and cleans existing directories if necessary.
\newline
2. BUILDING PACKAGES
\newline
* \texttt{PADE\_II}
\newline
* \texttt{DCS\_Regge}
\newline
* Utilities.
\newline
3. RUNNING \texttt{DCS\_Regge} FOR ALL INPUT FILES OF INTEREST
\newline
* Checks if the input file falls in the range of energies under investigation
\newline
* Runs \texttt{PADE\_II} with the current input file if it is in the considered range
\newline
* Runs \texttt{DCS\_Regge} if the current input file is in the considered range
\newline
* Calculates various resonance contributions  and subtracts them from the exact
scattering amplitudes in order to evaluate the non-resonance (direct) background.
\newline
4. OUTPUT DATA MANAGEMENT
\newline
* Stores the calculated data in the appropriate files.

\subsection {Using the code}

Using the code involves at least two steps.
\newline

\subsubsection {Step I}
In the parameter file \texttt{input/INPUT} (see Appendix A) one sets 
\newline
\texttt{is this the first run? :yes:}. The code evaluates the poles $P_i$ and the zeroes $Z_i$ of the \pd approximant (\ref{27}) for each collisional energy $E_k$, for the chosen set of files (see \#6-\#7 of Appendix A), in a region of the CAM plane, 
$ \texttt{x\_min} \le Re P_i, Z_i \le \texttt{x\_max}$, $\texttt{y\_min}\le Im P_i, Z_i \le \texttt{y\_max}$, with the values \texttt{x\_min}, \texttt{x\_max}, \texttt{y\_min}, and \texttt{y\_max} specified by the user in the file \texttt{input/INPUT} (see \#12 -\#15 in the Appendix A). 
\newline
One has the option of not including in the \pd approximant (\ref{27}) the {\it Froissart doublets}, i.e.,   pole-zero pairs, with the distance $|P_i-Z_j| <  \texttt{$\epsilon$}$, with the value of \texttt{$\epsilon$} set in \#11 of Appendix A. Such pairs often represent non-analytical noise existent in the input data \cite{G2}, and their removal may be beneficial.
\newline
Also, at this stage the program evaluates, for all energies, the exact DCS using Eq.(\ref{1}) or (\ref{3}). Provided the energy is entered in $meV$, and the reduced mass is in the 
$unified$ $atomic$ $mass$ $units$, ($u.a.u$ or $Daltons$) (see \#10 of Appendix A), the cross sections are in the units of angstroms squared
 (\AA$^2$).
 \newline
One then identifies Regge trajectories by plotting the pole positions vs. energy from the output file \texttt{output/dcs.pole}. (It is recommended to use the plot $Re P(E)$ vs. $E$, as the real parts of the pole positions are less sensitive to numerical noise).
In the plot the trajectories appear as continuous strings of poles, with additional poles scattered around them in a random manner. 
During the first run the program:

1) Writes the exact DCS (\ref{1}) vs. $\tr$ and $E$ into \texttt{output/dcs.dcs3d}, if \texttt{dcr3D}=1,
and, into \texttt{output/dcs.xdcs}, the exact DCS vs. $\tr$ for the last energy of the range.

2) Writes 
$|f^{NS}_{\nu^{\prime} \gets \nu}(\tr,E)|^2$, $|f^{FS}_{\nu^{\prime} \gets \nu}(\tr,E)|^2$ and $|f^{NS}_{\nu^{\prime} \gets \nu}(\tr,E)+ f^{FS}_{\nu^{\prime} \gets \nu}(\tr,E)|^2$ [cf. Eqs.(\ref{9}) -(\ref{10})
 into the file \texttt{output/dcs.nfdcs}, for the last energy in the range.

3) Writes $|S^{Pade} _\nn(E,J)|^2$ 
 vs. $E$ and $J$ into \texttt{output/dcs.prob3d}
  if {prob3D}=1 (cf. Eq.(\ref{27}).

4) Writes $|S^{Pade} _\nn(E,J)|$ and
$\R[S^{Pade} _\nn(E,J)]$ vs. $J$ into
\texttt{output/smprod} for the last energy in the range.
For comparison, the input values of $|S^J_\nn|$ and $\R[S^J _\nn]$ [cf. Eq.(\ref{3})] vs. integer $J$ are written into \texttt{output/inputvals}.

5) Writes the deflection function (\ref{22}) vs. $J$ and $E$ into \texttt{output/dcs.ph3d}, if \texttt{phs3D}=1.

6) Writes the deflection function (\ref{22}) and the phase $\Theta_\nn(\lm)$ in Eq.(\ref{18}) vs. $J$ into the file \texttt{output/phase}, for the last energy in the range.

7) Writes the first unfolded amplitude (\ref{12}), $|\tilde f(\va)|$ 
vs. $\va$, $\nl\times \pi \le \va \le \nr\times \pi$, and $E$ into \texttt{output/dcs.f3d}, if  
 \texttt{irun1}=1  
 and \texttt{dcr13D}=1. {(See \textcolor{black}{\#28-\#29} of Appendix A for the values of $\nl$ and $\nr$.)}  
 
 8) Writes the first unfolded amplitude (\ref{12}), $|\tilde f(\va)|$ and $\R[\tilde f(\va)]$,  vs. $\va$, $\nl\times \pi \le \va \le \nr\times \pi$, into the file \texttt{output/funf} for the last energy in the range, if  
 \texttt{irun1}=1. (If \texttt{irun1}=0, $\tilde f(\va)$ is not evaluated.) 
  
 9) Writes the second unfolded amplitude (\ref{13}), $|\tilde g(\va)|$, 
 vs. $\va$, $\nl\times \pi \le \va \le \nr\times \pi$, and $E$ into \texttt{output/dcs.g3d}, if  
 \texttt{irun1}=1 and \texttt{dcr13D}=1.
  
10) Writes the second unfolded amplitude (\ref{13}), $|\tilde g(\va)|$ and $\R[\tilde g(\va)]$ vs. $\va$, $\nl\times \pi \le \va \le \nr\times \pi$, into the file \texttt{output/gunf} for the last energy in the range, if    \texttt{irun1}=1. (If \texttt{irun1}=0, $\tilde g(\va)$ is not evaluated.)
  
11) Writes the NS amplitudes, $|f^{NS}_\nn(\tr,E|M)|^\np$ ($\np=1$ or $2$) in Eqs.(\ref{14}), for a chosen $\tr$
  and the specified ranges of $M$ and $E$, into \texttt{output/dcs.nsind}.
 
12) Writes the FS amplitudes, $|f^{FS}_\nn(\tr,E|M)|^\np$ {($\np=1$ or $2$)} in Eqs.(\ref{14}), for a specified $\tr$
  and the specified ranges of
    $M$ and $E$, into \texttt{output/dcs.fsind}.
 
13) For the specified range of energies, writes the exact sideway scattering amplitude, $|f_\nn(\tr,E)|^\np$ ($\np=1$ or $2$) in Eq.(\ref{1})
  and, for the specified range of $M$, $|\sum_M( f^{NS}_\nn(\tr,E|M)+f^{FS}_\nn(\tr,E|M)|^\np$ into \texttt{output/dcs.sw}.
  
14) Writes the forward scattering amplitudes in Eq.(\ref{16}), $|f^{FW}_\nn(E|M)|^\np$ {($\np=1$ or $2$)}, for the specified ranges of 
 $M$ and $E$, into \texttt{output/dcs.fwind}.
 
15) For the specified range of energies, writes the exact forward scattering amplitude, $|f_\nn(\tr=0,E)|^\np$ and, for the specified range of $M$, $|\sum_{M}f^{FW}_\nn(E|M)|^\np$ into \texttt{output/dcs.fw} {($\np=1$ or $2$)}.  
  
16) Writes the backward scattering amplitudes in Eq.(\ref{17}), $|f^{BW}_\nn(E|M)|^\np$ {($\np=1$ or $2$)}, for the specified ranges of 
$M$ and $E$ into \texttt{output/dcs.bwind}.
 
17) For the specified range of energies, writes the exact backward scattering amplitude, $|f_\nn(\tr=\pi,E)|^\np$ {($\np=1$ or $2$)} and, for the specified range of $M$, $|\sum_{M}f^{BW}_\nn(E|M)|^\np$ into \texttt{output/dcs.bw}.
  
18) Writes into \texttt{output/dcs.pole} the real and imaginary parts of the poles of the \pd approximant (\ref{27}) vs. $E$, within the limits specified 
  by the user in \#6-\#7 of the file \texttt{input/INPUT} (see Appendix A). The Froissart doublets can be excluded by choosing 
  a non-zero value of the threshold \#{11} of the file \texttt{input/INPUT}.
  
19) Writes into \texttt{output/dcs.zero} the real and imaginary parts of the zeroes of the \pd approximant (\ref{27}) vs. $E$, within the limits specifies 
  by the user in \#6-\#7 of the file \texttt{input/INPUT} (see Appendix A). 
  
\subsubsection {Step II } 
In the file \texttt{input/INPUT} one sets \texttt{is this the first run? :no:}. The code takes the first of the files in the energy range \texttt{E\_min} $\le E \le$ \texttt{E\_max} with \texttt{E\_min} and \texttt{E\_max} specified by the user in \#8-\#9 of the file \texttt{input/INPUT} (see Appendix A). It then displays all the poles at this energy within the specified range, from which the user chooses the one lying on the Regge trajectory of interest. 
 \newline
Next, if one sets
\texttt{follow trajectory by hand? :no:} the code will follow the trajectory automatically, choosing at the next energy the pole whose real part is closest to that of the pole chosen at the previous energy.
 \newline
If one chooses \texttt{follow trajectory by hand? :yes:} the program continues displaying the poles,
from which the user must choose the desired one for all values of energy by hand. 
(The "by hand" option is useful, e.g., when working with a poorly defined trajectory from {which} some of the poles may be missing.)
 \newline
In both cases, the program:

1) Writes the real and imaginary parts of the chosen pole's position [$n=1$, cf. Eq.(\ref{23})] 
 vs. $E$ into the file \texttt{output/dcs.traj}.

2) Writes the real and imaginary parts of the residue (\ref{28}) of the chosen pole vs. $E$ into the file \texttt{output/dcs.resid}.

3) For a specified $0<\tr<\pi$ writes into the file \texttt{output/dcs.swtind}
 $|f^{SW(tail)}_{\nu^{\prime} \gets \nu,1}(\tr,E)|^{\np}$ {($\np=1$ or $2$)}
as well  as 
{\color{black}$|f^{SW(tail)}_{\nu^{\prime} \gets \nu,1}(\tr,E|K)|^{\np}$ }in Eq.(\ref{24a}),
vs. $E$, for the chosen pole and  $1\le K \le \nr-1$, {where $\nr$ is specified by the user in \textcolor{black}{\#29} of the file \texttt{input/INPUT} (see Appendix A).}

4) Writes into the file \texttt{output/dcs.fwtind}  $|f^{FW(tail)}_{\nn,1} (E)|^{\np}$ {($\np=1$ or $2$)} in Eq.(\ref{25}), as well 
as $|f^{FW(tail)}_{\nn,1,\K} (E)|^{\np}$, vs. $E$, for $0 \le \K \le (\nr+1)/2$ (if $\nr$ is odd), or
$ \le \nr/2$ (if $\nr$ is even).

5) Writes into the file \texttt{output/dcs.bwtind}$|f^{BW(tail)}_{\nn,1}) (E)|^\np${($\np=1$ or $2$)} in Eq.(\ref{26}), as well 
as $|f^{BW(tail)}_{\nn,1,\K} (E)|^{\np}$,  vs. $E$, for $1 \le \K \le (\nr-1)/2$ (if $\nr$ is odd), or
$ \le \nr/2$ (if $\nr$ is even).

6) For a specified $0<\tr<\pi$ writes into the file \texttt{output/dcs.swsm}
{$|f^{SW(tail)}_{\nn,1}(\tr,E)|^{\np}$ ($\np=1$ or $2$)
and  $|f_{\nn} (\tr,E)-f^{SW(tail)}_{\nn,1}(\tr,E)|^{\np}$ vs. $E$ {[cf. Eq.(\ref{27a})].}

7) Writes into the file \texttt{output/dcs.fwsm}
$|f^{FW(tail)}_{\nn,1}(E)|^{\np}$ {($\np=1$ or $2$)}
and  
\newline
$|f_{\nn} (\tr=0,E)-f^{FW(tail)}_{\nn,1}(E)|^{\np}$ vs. $E$  {[cf. Eq.(\ref{25})].}

8) Writes into the file \texttt{output/dcs.bwsm}
$|f^{BW(tail)}_{\nn,1}(E)|^{\np}$ {($\np=1$ or $2$)}
and  
\newline
$|f_{\nn} (\tr=\pi,E)-f^{BW(tail)}_{\nn,1}(E)|^{\np}$ vs. $E$ [cf. Eq.(\ref{26})]. 

{9)} {\color{black}If a single energy is considered (the entry \#6 in the input file \texttt{INPUT} is set to 1)},
writes into the file \texttt{output/smof} $|\tilde f(\va)- \chi(\va-\pi) \tilde f^{(tail)}_1(\va)|^{\np}$ {($\np=1$ or $2$)}
 vs. $\va$,  
$\chi(x)\equiv 1$ for $x\ge 0$ and $0$ otherwise, [cf. Eqs.(\ref{24})].

{10)}  {\color{black}If a single energy is considered (the entry \#6 in the input file \texttt{INPUT} is set to 1)}, writes into the file \texttt{output/smog} $|\tilde g(\va)- \chi(\va-\pi) \tilde g^{(tail)}_1(\va)|^{\np}$ {($\np=1$ or $2$)}
vs. $\va$ [cf. Eqs.(\ref{24})].
\newline

Step II can be repeated several ($N$) times, thus making the program follow different Regge trajectories,
while choosing \texttt{E\_min} and \texttt{E\_max} as is convenient. If $N$ different Regge trajectories were 
followed in $N>1$ runs, the output files will store the data:
\newline
File \texttt{output/dcs.swsm}: 
\newline
$|\sum_{n=1}^Nf^{SW(tail)}_{\nn,n}(\tr,E)|^{\np}$ and
\newline
 $|f_{\nn} (\tr,E)-\sum_{n=1}^N f^{SW(tail)}_{\nn,n}(\tr,E)|^{\np}$. 
 \newline
File \texttt{output/dcs.fwsm}: 
\newline
$|\sum_{n=1}^Nf^{FW(tail)}_{\nn,n}(E)|^{\np}$ and 
\newline
$|f_{\nn} (\tr=0,E)-\sum_{n=1}^N f^{FW(tail)}_{\nn,n}(E)|^{\np}$.
\newline
File \texttt{output/dcs.bwsm}: 
\newline
$|\sum_{n=1}^Nf^{BW(tail)}_{\nn,n}(E)|^{\np}$ and 
\newline
$|f_{\nn} (\tr=\pi,E)-\sum_{n=1}^N f^{BW(tail)}_{\nn,n}(E)|^{\np}$.
\newline
File {\texttt{output/smof}:
\newline
$|\tilde f(\va)- \chi(\va-\pi)\sum_{n=1}^N \tilde f^{(tail)}_n(\va)|^{\np}$.
\newline
File {\texttt{output/smog}: 
\newline
$|\tilde g(\va)- \chi(\va-\pi)\sum_{n=1}^N \tilde g^{(tail)}_n(\va)|^{\np}$.

Auxiliary output files, not described explicitly in this section, are required for repeating Step II and can be removed from the \texttt{DCS/output} directory by running the script \texttt{DCS/clean\_aux} at the end of the calculation.
 
\section{Examples of using \texttt{DCS\_Regge} }
Three examples of using the $\texttt{DCS\_Regge}$ package are provided. 
These are essentially the same systems which were used as examples in \cite{ICS_REGGE}, 
with the difference that now we analyse the differential rather than integral cross section. 
The input files for these examples are included in the package.
\subsection{Example 1: The hard sphere model (Regge trajectory of the type I)}
The first example 
involves the $S$-matrix element  for potential (single channel) scattering off a hard sphere of a radius $R-d$ surrounded by a thin semi-transparent layer of a radius $R$. Since the problem is essentially a single-channel one, we omit the subscripts $\nn$.)
The spherically symmetric potential $V(r)$ in Eq.(\ref{4a}) is infinite for $r<R-d$, has a rectangular well of a depth $V$ for $R-d<r<R$, a zero range barrier $\Omega \delta(r-R)$ ($\delta(x)$ is the Dirac delta), and vanishes elsewhere \cite{PLA}. In this example the energy of a non-relativistic particle of a mass $\mu=1$ $u.a.u$ varies from $1$ $meV$ to $100$ $meV$, the radii of the hard sphere and the width of the well $d$ are $2.045$\AA\ and $0.592$\AA, respectively, $V=165$ $meV$, and
 $\Omega = 1.023$ $meV\cdot$ \AA. In this range, there is a single resonance Regge trajectory, originating at $J=0$ in the bound state of the well at about $- 14$ $meV$. 
A detailed discussion of this model can be found in Refs.\cite{PLA} where the Regge trajectories were obtained by direct integration of the radial Schroedinger equation for complex values of  $J$.
The only difference with \cite{ICS_REGGE} is that, in order to mimic reactivity, we introduce a Gaussian factor [cf. Eq.(\ref{4b})]
with $\Delta J =5$, so that the PWS (\ref{3}) converges after $N_{max} \approx 10$ terms.
Here we seek to repeat the results of \cite{PLA} by evaluating the $S$ matrix element for integer $J$'s, and then using the \pd reconstruction.
The data files are in the directory  \texttt{input/BOUND.DCS}.  The corresponding DCS is shown in \autoref{plot:Fig1}a).
\newline 
\underline {\it Step I } 
\newline
In the directory  \texttt{input}, copy the file  \texttt{INPUT.BOUND.DCS.30} into the file  \texttt{INPUT}.
Run the code. Use the files \texttt{dcs.xdcs} 
and \texttt{dcs.nfdcs} to plot the exact DCS and its NS and FS   components [cf. Eqs.(\ref{9})-(\ref{10})]
at $kR=30$, where $k$ is the particle's wave vector. 
Use the file  \texttt{phase} to plot the deflection function (\ref{22}). Use the file \texttt{funf} to plot the {\color{black}first} unfolded amplitude
$\tilde f(\va)$ in Eq.(\ref{12}).
\newline 
\underline {\it Step II }
In the file \texttt{INPUT} change \texttt{is this the first run? :yes:} to  \texttt{is this the first run? :no:}.
When prompted, select the pole with $ReJ \approx 8.2356$ and $Im J \approx 0.3350$.
Run the code. Use the file  \texttt{output/smof} to plot  the difference between $\tilde f(\va)$ and the exponential tail 
$\tilde f^{tail}$ in Eqs.(\ref{24}).
\newline 
\underline {\it Step III }
\newline
In the directory  \texttt{input}, copy the file  \texttt{INPUT.BOUND.DCS} into the file  \texttt{INPUT}.
Run the code to completion.  Use the file  \texttt{output/dcs.pole} to plot real and imaginary parts of the poles vs. energy, and identify the relevant Regge trajectory. Use the file  \texttt{output/dcs.sw} to plot the DCS at $\tr=75^o$ in the specified range of energies.
Use the files  \texttt{output/dcs.nsind} and \texttt{output/dcs.fsind}  to plot
$f^{NS}(\tr,E|M)$, $M=0,1,2$ and $f^{FS}(\tr,E|M)$, $M=0,1$ in Eqs.(\ref{14}).
\newline 
\underline {\it Step IV }
In the file \texttt{INPUT} change \texttt{is this the first run? :yes:} to  \texttt{is this the first run? :no:}. Run the code. 
When prompted, select the pole with $ReJ \approx 5.954$ and $Im J \approx 7.67E-02$. Finish the calculation. 
Use the file  \texttt{output/dcs.traj} to plot the Regge trajectory of the chosen pole.
Use the file  \texttt{output/dcs.swsm} to plot the resonance contribution 
$f^{SW(tail)}(\tr=75^o,E)$ in Eq.(\ref{27a})
and the difference $f(\tr=75^o,E)-f^{SW(tail)}(\tr=75^o,E)$.
\newline
The correct results, obtained in steps $I$-$IV$, are shown in \autoref{plot:Fig7}.
\newline
\begin{figure}[h] 
\subfloat{\includegraphics[angle=0,width=11.3 cm]{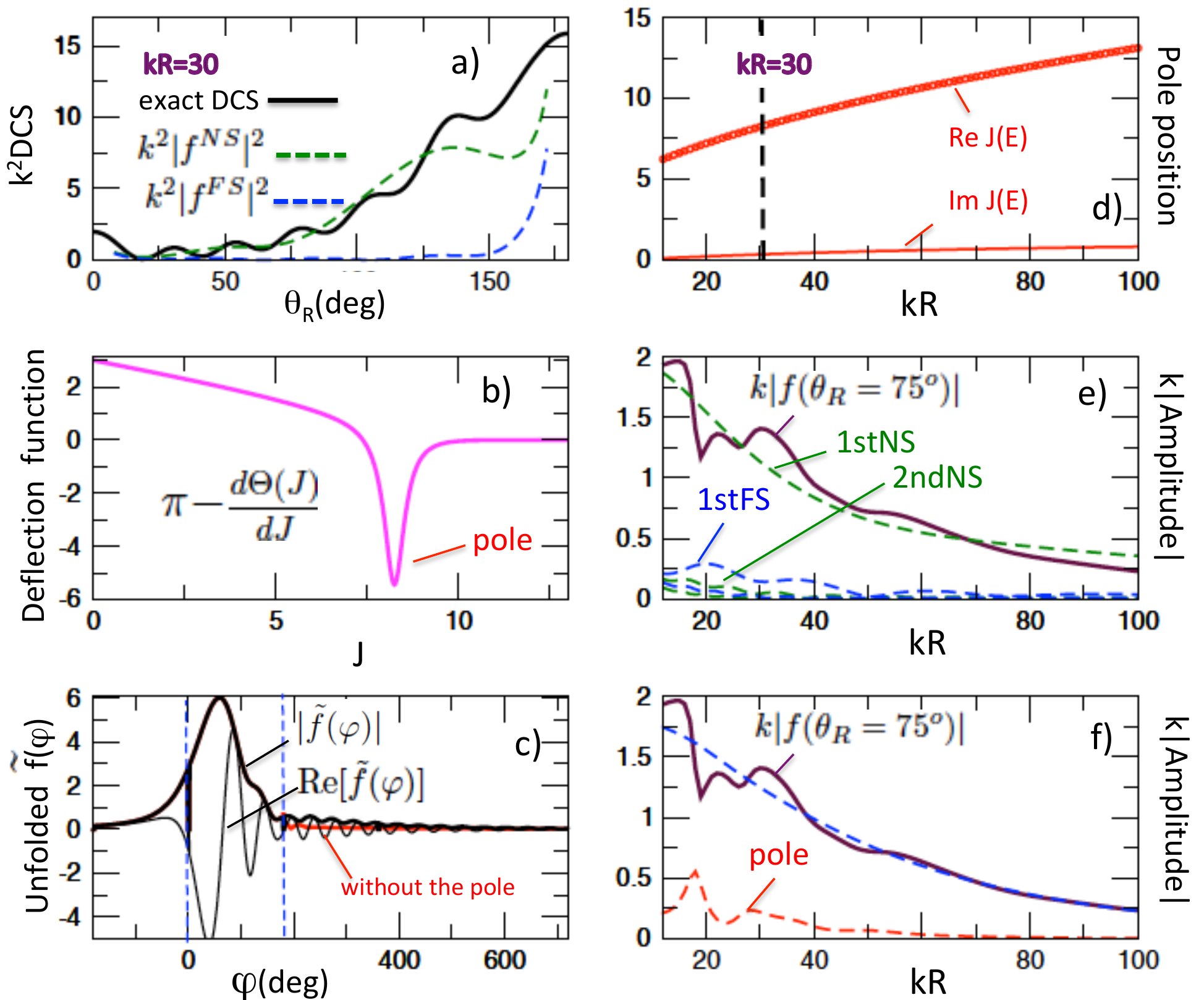}}
\caption
{ {\bf Example 1: the hard-sphere model (type I).}
\newline
\underline{\bf The task:} to analyse the behaviour of the DCS in \autoref{plot:Fig1}a) as a function of energy, at $\theta _R=75^o$. 
\newline
{\bf (a)} The full DCS (solid) and its NS and FS  components (\ref{9}) - (\ref{10}) at $kR=30$.
{\bf {\it Thus,}} the presence of both components suggests that the oscillatory behaviour 
of the DCS is a NS-FS interference effect. 
\newline
{\bf (b)}  The deflection function (\ref{22}) at $kR=30$ exhibits a narrow dip.
{\bf {\it Thus,}} this suggests the presence of a single resonance Regge pole.
The particle may be trapped in a metastable intermediate state at the total angular momentum $J\approx 7$.
\newline
{\bf (c)}  The unfolded amplitude $\tilde f(\va)$ at $kR=30$ has a small but slowly decreasing component, which extends beyond the first NS zone, $0<\va <\pi$. Also shown (in red) 
is  the result 
of subtracting from $\tilde g(\va)$ the "exponential tail", $\tilde g^{tail} (\va)$ associated with the Regge pole presented in (d).
 {\it Thus,} for $\va> \pi$, one has an exponential decay of a single metastable state. This process, 
 resulting in scattering into the first FS zone,  is likely to interfere with scattering into the first NS zone.
\newline
{\bf (d)}  Real (circles) and imaginary (solid) parts of the Regge pole positions as functions of the particle's energy. 
 {\it Thus,} a single resonance affects the angular scattering shown in \autoref{plot:Fig1}a) in the energy range $10\le kR \lesssim60$.
 \newline
{\bf (e)} The sideway scattering amplitude at $\tr=75^o$ vs. energy. Also shown by the dashed lines are 
the leading $NS$ and $FS$ contributions in Eqs.(\ref{14}), $f^{NS}(\tr=75^o,E|M)$ and $f^{FS}(\tr=75^o,E|M)$.
 {\it Thus,} the probable cause of the  oscillations in the DCS  at $\tr=75^o$ for $10 \le kR\lesssim60$ 
 is the interference between the decays into various NS and FS zones.
  \newline
{\bf (f)} Contribution of the resonance to scattering at $\tr=75^o$, as given by $f^{SW(tail)}(\tr)$ in Eq.(\ref{27a}) (red dashed). 
Also shown (blue dashed) is the result of subtracting it from the exact scattering amplitude, $f(\tr=75^o)-f^{SW(tail)}(\tr=75^o)$. 
\newline
 \underline{{\bf Conclusion:}} Sideway scattering cross section at $\tr=75^o$ in the range $10 \le kR\lesssim60$ is shaped by the interference between scattering into the first NS zone [the first equation in (\ref{14})], and what the resonance contributes beyond the first NS zone  [the first equation in Eq.(\ref{24})].
  At lower energies, 
 where the lifetime of the resonance is large, the latter contribution is itself structured due to interference between multiple 
 rotations of the complex. 
 }\label{plot:Fig7}
\end{figure}
\subsection{Example 2: The hard sphere model (Regge trajectory of the type II)}
This is the same model as in Example 1, but  with  $\Omega = 66.463$ $meV\cdot$ \AA, considered in the range of collision energies from $40$ $meV$ to $100$ $meV$. In this case, there is a single resonance Regge trajectory, originating at $J=0$ in a metastable state with the real part of about $48$ $meV$. 
The data files are in the directory  \texttt{input/META.DCS}. The corresponding DCS is shown in \autoref{plot:Fig1}b).
\newline 
\underline {\it Step I } 
\newline
In the directory  \texttt{input}, copy the file  \texttt{INPUT.META.DCS.60} into the file  \texttt{INPUT}.
Run the code. Use the files \texttt{dcs.xdcs} and \texttt{dcs.nfdcs} to plot the exact DCS and its NS and FS  components [cf. Eqs.(\ref{9})-(\ref{10})] 
at $kR=60$, where $k$ is the particle's wave vector. Use the file  \texttt{phase} to plot the deflection function (\ref{22}). Use the file \texttt{gunf} to plot the {\color{black} second} unfolded amplitude
$\tilde g(\va)$ in Eq.(\ref{13}).
\newline 
\underline {\it Step II }
In the file \texttt{INPUT} change \texttt{is this the first run? :yes:} to  \texttt{is this the first run? :no:}.
When prompted, select the pole with $ReJ \approx 3.778$ and $Im J \approx 0.5097$.
Run the code. Use the file  \texttt{output/smog} to plot  the difference between $\tilde g(\va)$ and the exponential tail 
{\color{black}$\tilde g^{(tail)}_1(\va)$} in Eqs.(\ref{24}).
\newline 
\underline {\it Step III }
\newline
In the directory  \texttt{input}, copy the file  \texttt{INPUT.META.DCS} into the file  \texttt{INPUT}.
Run the code to completion.  Use the file  \texttt{output/dcs.pole} to plot real and imaginary parts of the poles vs. energy, and identify the relevant Regge trajectory. Use the file  \texttt{output/dcs.bw} to plot the backward DCS at $\tr=180^o$ in the specified range of energies.
Use the file \texttt{output/dcs.bwind}  to plot
{\color{black}$f^{BW}(E|M)$}, $M=1,2,3$ in Eqs.(\ref{17}).
\newline 
\underline {\it Step IV }
In the file \texttt{INPUT} change \texttt{is this the first run? :yes:} to  \texttt{is this the first run? :no:}. Run the code.
When prompted, select the pole with $ReJ \approx 0.1479$ and $Im J \approx 3.6021$. Finish the calculation.
 Use the file  \texttt{output/dcs.traj} to plot the Regge trajectory of the chosen pole.
 Use the file  \texttt{output/dcs.bwsm} to plot the two terms in Eq.(\ref{26}).
\newline
The correct results, obtained in steps $I$-$IV$, are shown in \autoref{plot:Fig8}.
\subsection{Example 3: The $F+H_2(v=0,j= 0,\Omega=0) \to H F(v'=2, j'=0,\Omega'=0)+H$ reaction.  (Two pseudo-crossing Regge trajectories)}
This example uses realistic numerical data obtained in Ref. \cite{DARIO04}, and analysed previously  in Refs. \cite{FH3}-\cite {PCCP}.
In the collision energy range $20-50$ $meV$  there are two resonance Regge trajectories, labelled $I$ and $II$ (we follow the notations of \cite{PCCP}). (For the effects in the state-to-state integral cross section 
see \cite{ICS_REGGE}). At the collision energy of about $38$ $meV$ the imaginary parts of the trajectories cross, while the real parts do not (for details see Ref.\cite{FH3}).
The data files are in the directory 
\texttt{input/FH2.DCS}. The corresponding DCS is shown in \autoref{plot:Fig1}c).
\newline 
\underline {\it Step I } 
\newline
In the directory  \texttt{input}, copy the file  \texttt{INPUT.FH2.DCS.32.27} into the file  \texttt{INPUT}.
Run the code. Use the files \texttt{dcs.xdcs} and \texttt{dcs.nfdcs} to plot the exact DCS and its NS and FS   components [cf. Eqs.(\ref{9})-(\ref{10})]
at $E=32.27$ $meV$. Use the file  \texttt{phase} to plot the deflection function (\ref{22}). Use the file \texttt{gunf} to plot the {\color{black} second} unfolded amplitude
$\tilde g(\va)$ in Eq.(\ref{13})).
\newline 
\underline {\it Step II }
In the file \texttt{INPUT} change \texttt{is this the first run? :yes:} to  \texttt{is this the first run? :no:}. Run the code. 
When prompted, select the resonance pole ($II$) with $ReJ \approx 4.911$ and $Im J \approx 0.602$.
Use the file  \texttt{output/smog} to plot  the difference between $\tilde g(\va)$ and the exponential tail,  $\tilde g^{(tail)}_{II}(\va)$ in Eq.(\ref{24}), corresponding to the chosen pole.
Run the code again.
When prompted, select the other pole ($I$) with $ReJ \approx 6.676$ and $Im J \approx 1.05$.
Use the file  \texttt{output/smog} to plot  the difference between $\tilde g(\va)$ and the sum of both exponential tails in Eq.(\ref{24}), 
 $\tilde g^{(tail)}_I(\va)+\tilde g^{(tail)}_{II}(\va)$.
\newline 
\underline {\it Step III }
\newline
In the directory  \texttt{input}, copy the file  \texttt{INPUT.FH2.DCS} into the file  \texttt{INPUT}.
Run the code to completion.  Use the file  \texttt{output/dcs.pole} to plot real and imaginary parts of the poles vs. energy, and identify the relevant Regge trajectories. Use the file  \texttt{output/dcs.fw} to plot the forward DCS at $\tr=0^o$ in the specified range of energies.
Use the file \texttt{output/dcd.fwind}  to plot
$f^{FW}_{\nu'\gets \nu}(E|M)$, $M=1,2,3$ in Eqs.(\ref{16}).
\newline 
\underline {\it Step IV }
In the file \texttt{INPUT} change \texttt{is this the first run? :yes:} to  \texttt{is this the first run? :no:}. Run the code.
When prompted, select the pole with $ReJ \approx 1.603$ and $Im J \approx 1.103$. Finish the calculation.
 Use the file  \texttt{output/dcs.traj} to plot the Regge trajectory of the pole ($II$). Use the file  \texttt{output/dcs.fwtind} to plot the contributions (\ref{25a}) of the pole ($II$) to the forward scattering amplitude.
Run the code again. When prompted, select the pole with $ReJ \approx 5.590$ and $Im J \approx 1.410$.
Use the file  \texttt{output/dcs.traj} to plot the Regge trajectory of the pole ($I$).
Use the file  \texttt{output/dcs.fwtind} to plot the contributions (\ref{25a}) of the pole ($I$) to the forward scattering amplitude.
Use the file \texttt{output/dcs.fwsm} to plot the coherent sum of two pole contributions.
\newline
 To plot the trajectory of the Regge zero shown in green in \autoref{plot:Fig9}d
change in the  \texttt{INPUT} file parameters $\#2$ to \texttt{:yes:}, $\#6$ to $63$, 
$\#12$ to $3$, $\#13$ to $10$, and $\#15$ to $0.27$ and run the code again. The trajectory is in the file  \texttt{output/dcs.zero}.
These changes
 will help to get rid of irrelevant zeros, such as those belonging to the Froissart doublets \cite{G3}.
 \newline
 \textcolor{black}{The correct results 
 are shown in \autoref{plot:Fig9}. }
\begin{figure}[ht]
\subfloat{\includegraphics[angle=0,width=11.3cm]{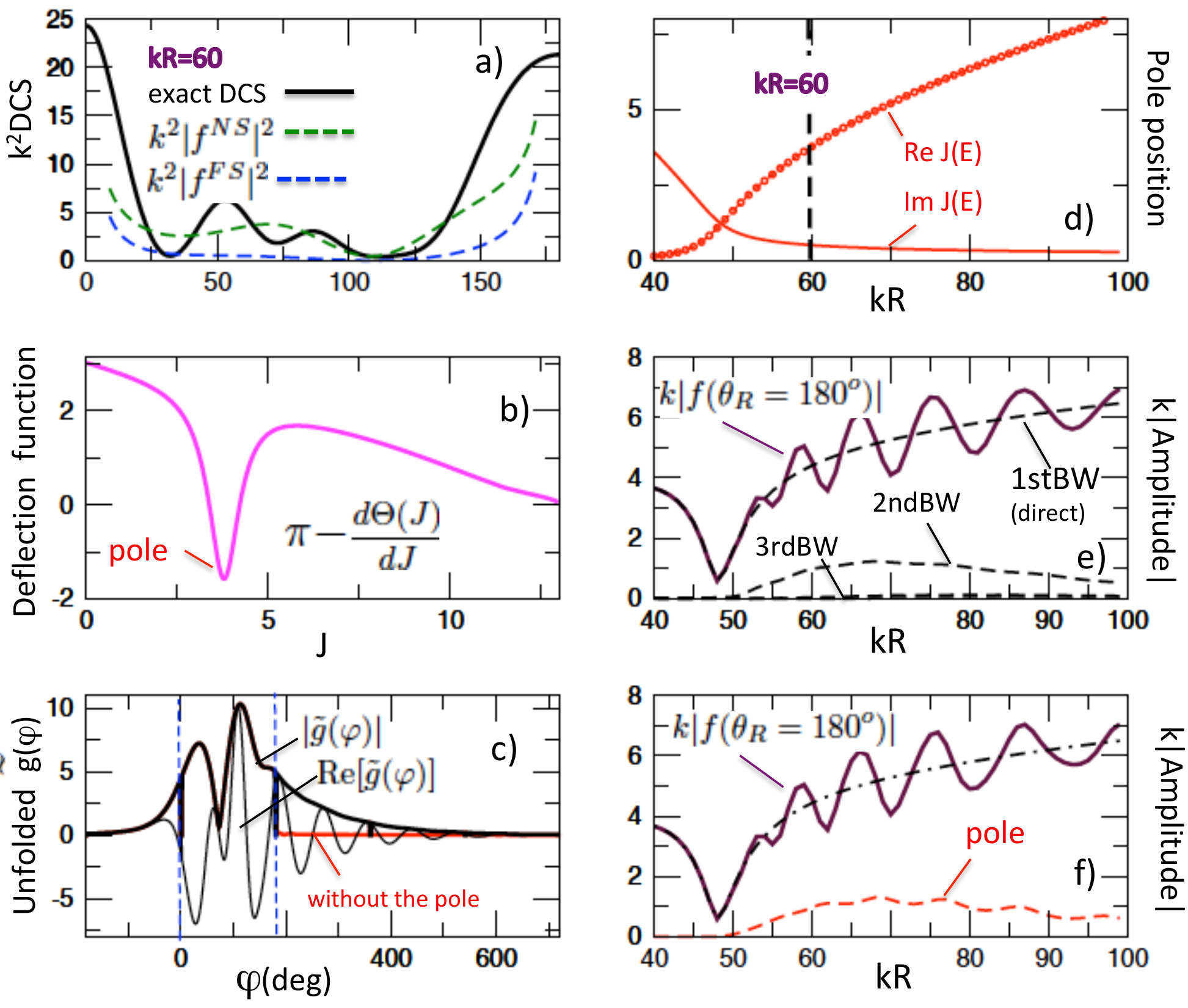}}
\caption
{ {\bf Example 2: the hard-sphere model (type II).}
\newline
\underline{\bf The task:} to determine the origin of the backward scattering oscillations in \autoref{plot:Fig1}b). 
\newline
{\bf (a)} The full DCS (solid) and its NS and FS  components (\ref{9})-(\ref{10}) at $kR=60$.
{\bf {\it Thus,}} the presence of both components suggests that the oscillatory behaviour 
of the DCS is a NS-FS interference effect. 
\newline
{\bf (b)}  The deflection function (\ref{22}) at $kR=60$ exhibits a narrow dip.
{\bf {\it Thus,}} this suggests the presence of a single resonance Regge pole.
The particle may be trapped in a metastable intermediate state at the total angular momentum $J\approx 4$.
\newline
{\bf (c)}  The unfolded amplitude $\tilde g(\va)$ at $kR=60$ extends beyond the first NS zone, $0<\va <\pi$
into the first FS and the second NS zone. Also shown (in red) is the result 
of subtracting from $\tilde g(\va)$ the "exponential tail", $\tilde g^{tail} (\va)$ associated with the Regge pole [cf. {\bf(d)}].
  {\it Thus,} for $\va> \pi$, one has an exponential decay of a single metastable state. 
The decay rate is such that the trapped particle can escape into the backward direction after one full rotation. This process is likely to interfere
 with the direct backward scattering resulting from a "head-on collision" at $J\approx 0$.
\newline
{\bf (d)}  Real (circles) and imaginary (solid) parts of the Regge pole positions as functions of the particle's energy. 
 {\it Thus,} a single resonance affects the scattering shown in \autoref{plot:Fig1}b) in the energy range $55\lesssim kR \le100$. 
 \newline
{\bf (e)} The backward scattering amplitude vs. energy. Also shown (dashed lines) are the backward scattering sub-amplitudes
in Eq.(\ref{17}),  $f_\nn(\tr=\pi,E|M)$, $M=0,1,2$.
 {\it Thus,} the probable cause of the  oscillations in the DCS  at $\tr=\pi$ for $kR>55$ 
 is the interference between  the scenarios where the particle bounces back immediately, or completes a single full 
 rotation around the potential's core. For $kR < 55$, the resonance is short-lived, and the second scenario cannot be realised.  
  \newline
{\bf (f)} Contribution of the resonance in {\bf(d)}  to backward scattering, as given by the second term in Eq.(\ref{26}) (dashed). 
Also shown by a dot-dashed line is the result of subtracting it from the $f_\nn(\tr=\pi)$. 
\newline
 \underline{{\bf Conclusion:}} As  expected, the oscillations in the backward DCS [cf.{\bf(e),(f)}] result from the interference between the rapid direct recoil and the decay of a trapped particle, after completing one full rotation.}\label{plot:Fig8}
\end{figure}
\begin{figure}[h]
\subfloat{\includegraphics[angle=0,width=11.3cm]{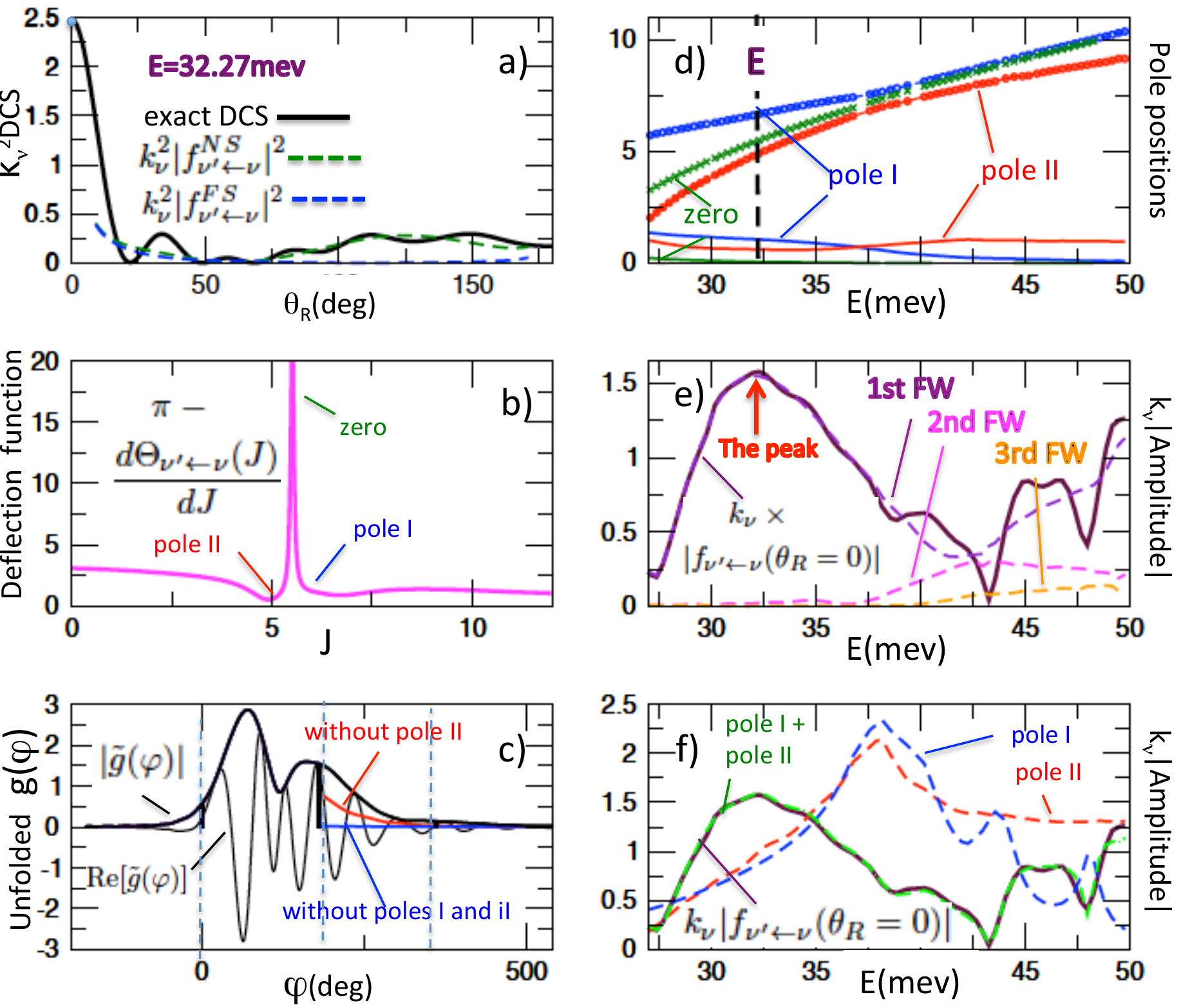}}
\caption
{ {\bf Example 3: the $F+H_2(0,0,0)\to HF(2,0,0)+H$ reaction.}
\newline
\underline{\bf The task:} to determine the origin of the forward scattering peak in \autoref{plot:Fig1}c). 
\newline
{\bf (a)} The full DCS (solid) and its NS and FS  components (\ref{9})-(\ref{10}) ( dashed) at the collision energy $E=32.27$ $meV$.
{\bf {\it Thus,}} the oscillatory pattern is likely to be a nearside-farside effect.
\newline
{\bf (b)}  The deflection function (\ref{22}) at $E=32.27$ $meV$ exhibits two dips separated by a sharp peak.
{\bf {\it Thus,}} resonance effects are likely to play an important role.
\newline
{\bf (c)}  The unfolded amplitude $\tilde g(\va)$ at $E=32.27$ $meV$ extends beyond the first NS zone, $\va > \pi$. Also shown (in red and blue) the results 
of subtracting from $\tilde g(\va)$ the "exponential tails", $\tilde g^{(tail)}_{II} (\va)$ and $\tilde g^{(tail)}_I (\va)$, in (\ref{24}), associated with the two Regge poles, 
labelled $I$ and $II$ 
in ${\bf (d)}$ (we follow the notations of \cite{PCCP}).
 {\it Thus,} forward scattering at $\tr=0$ occurs mostly via capture into intermediate states. There is a pair of such states involved. 
\newline
{\bf (d)}  Real (circles) and imaginary (solid) parts of the Regge pole positions as functions of collision energy. 
Also shown (in green) are the real (crosses) and imaginary (solid) parts of the position of a Regge zero, responsible for the sharp 
peak in the deflection function in ${\bf (b)}$. 
 {\it Thus,} both resonances are likely to affect the DCS in the entire energy range $27-50$ $meV$. The first state ($I$)
 becomes ever more long-lived as the energy increases, while the opposite happens to the state $II$. 
 \newline
{\bf (e)} The forward scattering amplitude (solid)  as a function of collision energy. Also shown by the dashed lines are the forward scattering sub-amplitudes
in Eq.(\ref{16}),   {\color{black}$f^{FW}_{\nn} (E|M)$}, $M=0,1,2$.
 {\it Thus,} the structure in the DCD  at $\tr=0$ for $E>37$ $meV$ is likely to result from the formation of an intermediate triatomic, 
 which can return to the forward direction after up to two complete rotations.
  \newline
{\bf (f)} The  {\color{black} exact} forward scattering amplitude (solid), and the contributions of the two resonances ${\bf (d)}$, shown by dashed lines, as  given by Eq.(\ref{25a}). 
Also shown (green, dashed) is the coherent sum of the two contributions. 
\newline
\underline{\bf Conclusion:}  The forward scattering peak at $E\approx 32-33$ $meV$ is caused by constructive interference 
 between the decays of the resonances $I$ and $II$ into the first forward zone \cite{PCCP}.}\label{plot:Fig9} 
\end{figure} 
\section{Summary}
In summary, we present a user friendly computer code which evaluates the contribution a resonance Regge trajectory makes to a reactive differential  cross section. Regge poles positions and residues are calculated using numerical values of the corresponding scattering matrix element by Pad\'{e}\q reconstruction. 
The code can be used for analysing reactive transitions with  zero initial and final helicity quantum numbers. 
\section{Acknowledgements:}
We acknowledge support of the Basque Government (BERC 2018e2021 Program, \textcolor{black} {Grant No. IT986-16,} and ELKARTEK grants KK-2020/00049, KK-2020/00008), and MINECO, the Ministry of Science and Innovation of Spain (BCAM Severo Ochoa accreditation SEV-2017-0718 and Grants \textcolor{black} {PGC2018-101355-B-100(MCIU/AEI/FEDER,UE}, PID2019-104927GB-C22). 
This work has been possible thanks to the support of the computing infrastructure of the i2BASQUE academic network, DIPC Computer Center and the technical and human support provided by IZO-
SGI SGIker of UPV/EHU and European funding (ERDF and ESF).
\section{Appendix A: Example of  \texttt{DCS\_Regge} parameter file \texttt{INPUT}} 
\begin{figure}[ht]
\begin{center}
\subfloat{\includegraphics[angle=0,width=11.8 cm]{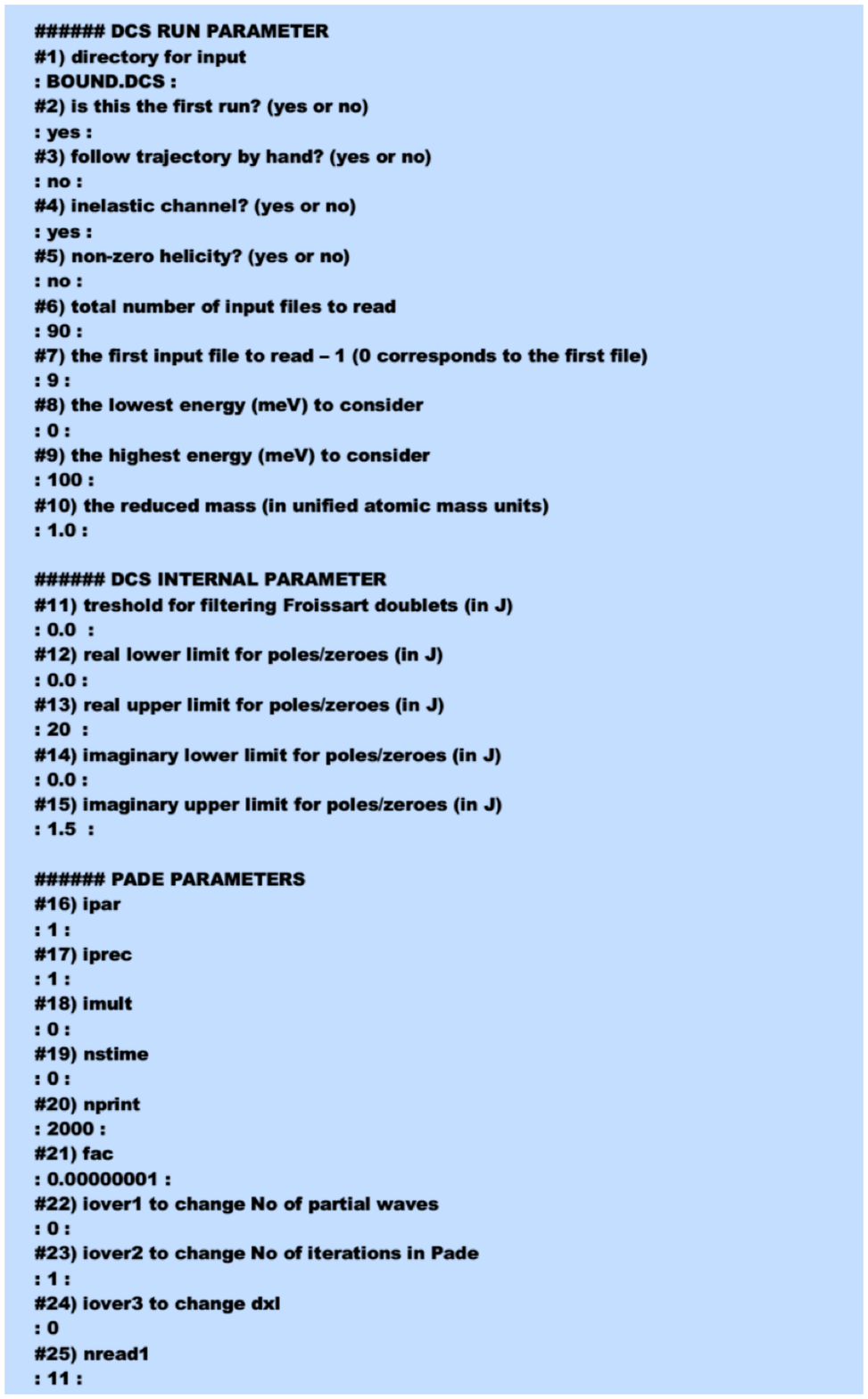}}
\end{center}
\caption{An example of the \texttt{INPUT} file.} \label{plot:Fig10}
\end{figure}
\begin{figure}[ht]
\begin{center}
\subfloat{\includegraphics[angle=0,width=13.5 cm]{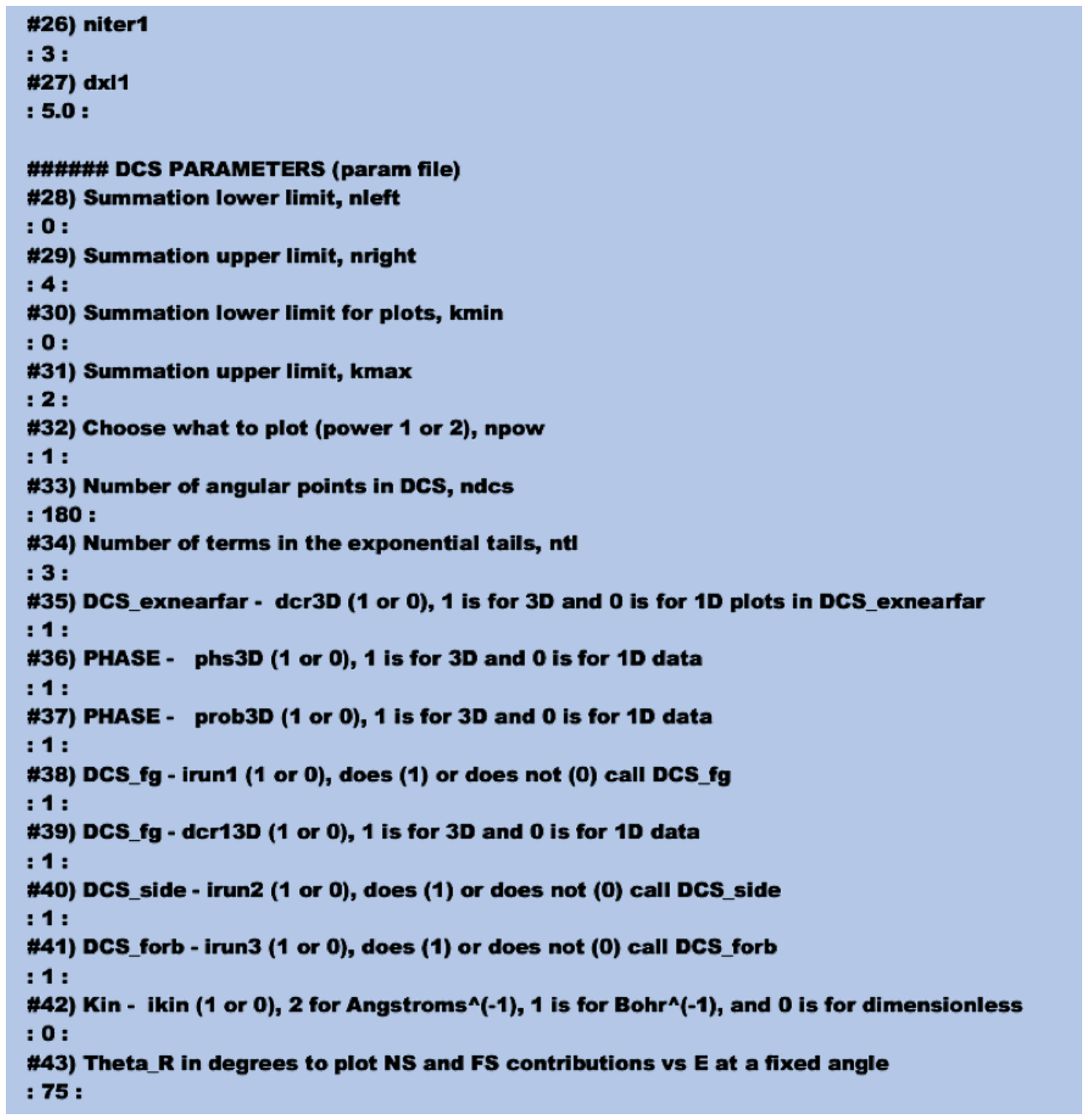}}
\end{center}
\caption{An example of the \texttt{INPUT} file (continued).}\label{plot:Fig11}
\end{figure}
See \autoref{plot:Fig10} and \autoref{plot:Fig11}.
\section{Appendix B: Example of  \texttt{DCS\_Regge} input file  \texttt{20}} 
\begin{figure}[ht]
\begin{center}
\subfloat{\includegraphics[angle=0,width=12 cm]{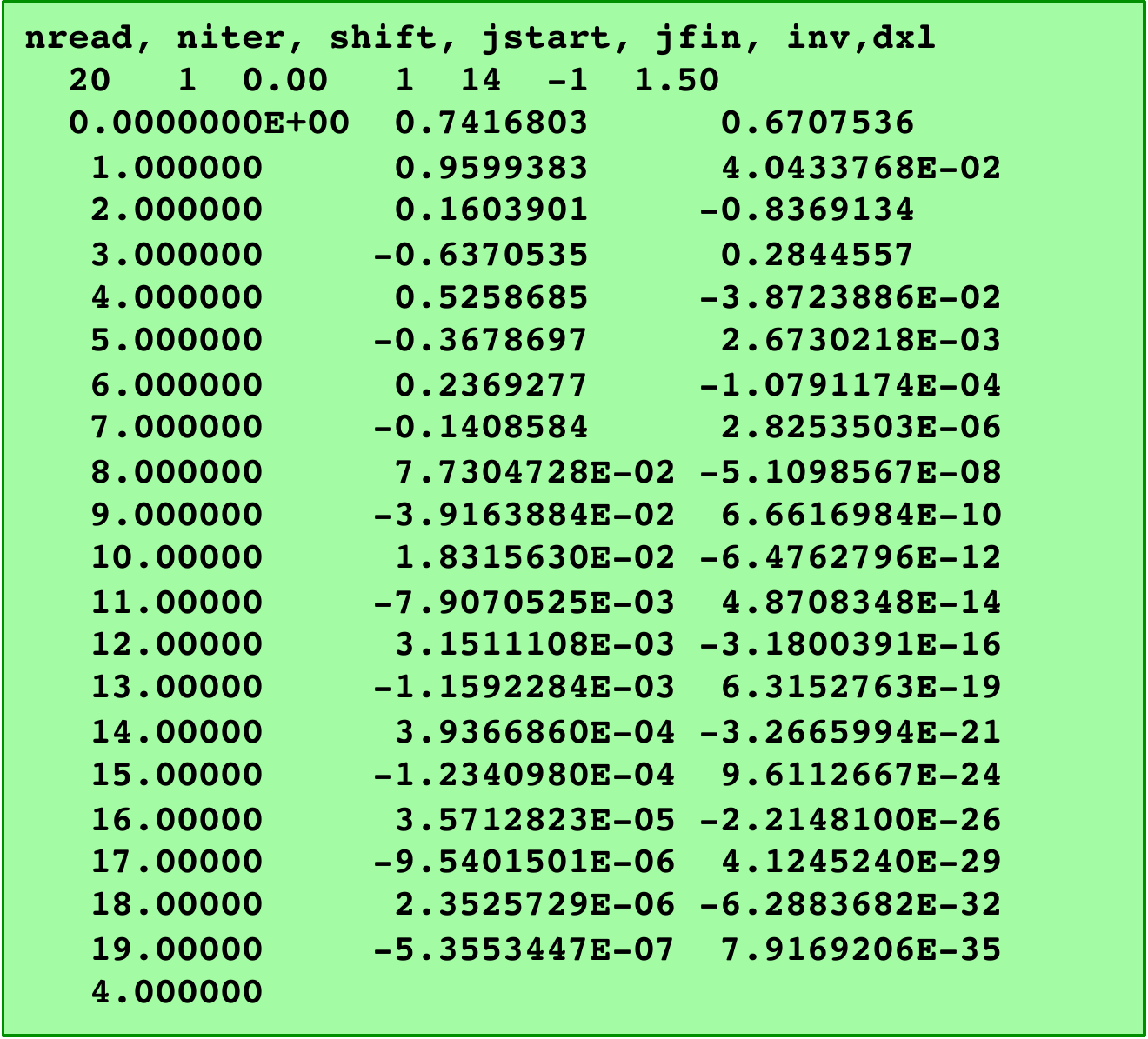}}
\end{center}
\caption{The input file \texttt{20}, which contains the input parameters required to run  \texttt{\PD\_II}, the values of the $S$-matrix element for different values of $J$, and the value of collision energy in $meV$. }\label{plot:Fig12}.
\end{figure}
The contents of the file, also described in \cite{PADE2}, include:
\newline
\x {nread}: the number of partial waves read,
\newline
\x{niter}: the number of iterations to remove the quadratic phase,
\newline
\x{sht}: this shifts the input grid points and may be used to avoid exponentiation of extremely large number when evaluating the polynomials involved. The value \x {sht} = \texttt{nread}$/2$ is suggested for a large number of partial waves. 
\newline
\x{jstart} and  \x{jfin}: with all input points numbered by \texttt{j} between 1 and N, determine a range \texttt{jstart} $\le j \le$  \texttt{jfin} to be used for the \pd reconstruction.
\newline
\x{inv}: set to $-1$, not used in present calculations, 
\newline
\x{dxl}: determines the width of the strip in which poles and zeroes are removed while evaluating the quadratic phase in Eq. (\ref{9}).
See \autoref{plot:Fig12}.
\section{Appendix C: Structure of the \texttt{DCS} directory}
\begin{figure}[ht]
\subfloat{\includegraphics[angle=0,width=15 cm]{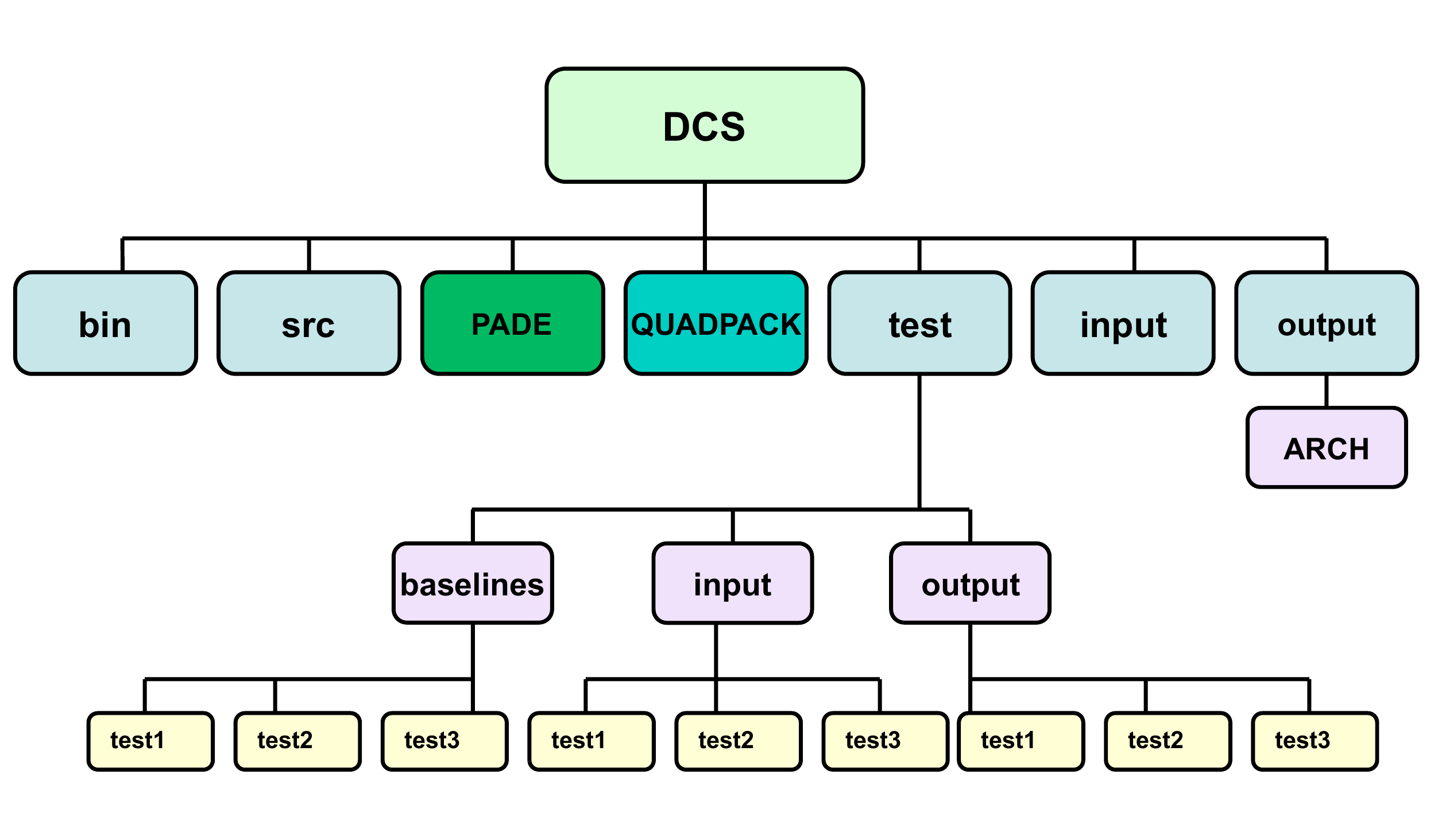}}
\caption{Detailed structure of the  \texttt{DCS} directory.}\label{plot:Fig13}
\end{figure}
See \autoref{plot:Fig13}.
\section{Appendix D: Structure of the \texttt{PADE} directory}
\begin{figure}[ht]
\subfloat{\includegraphics[angle=0,width=15 cm]{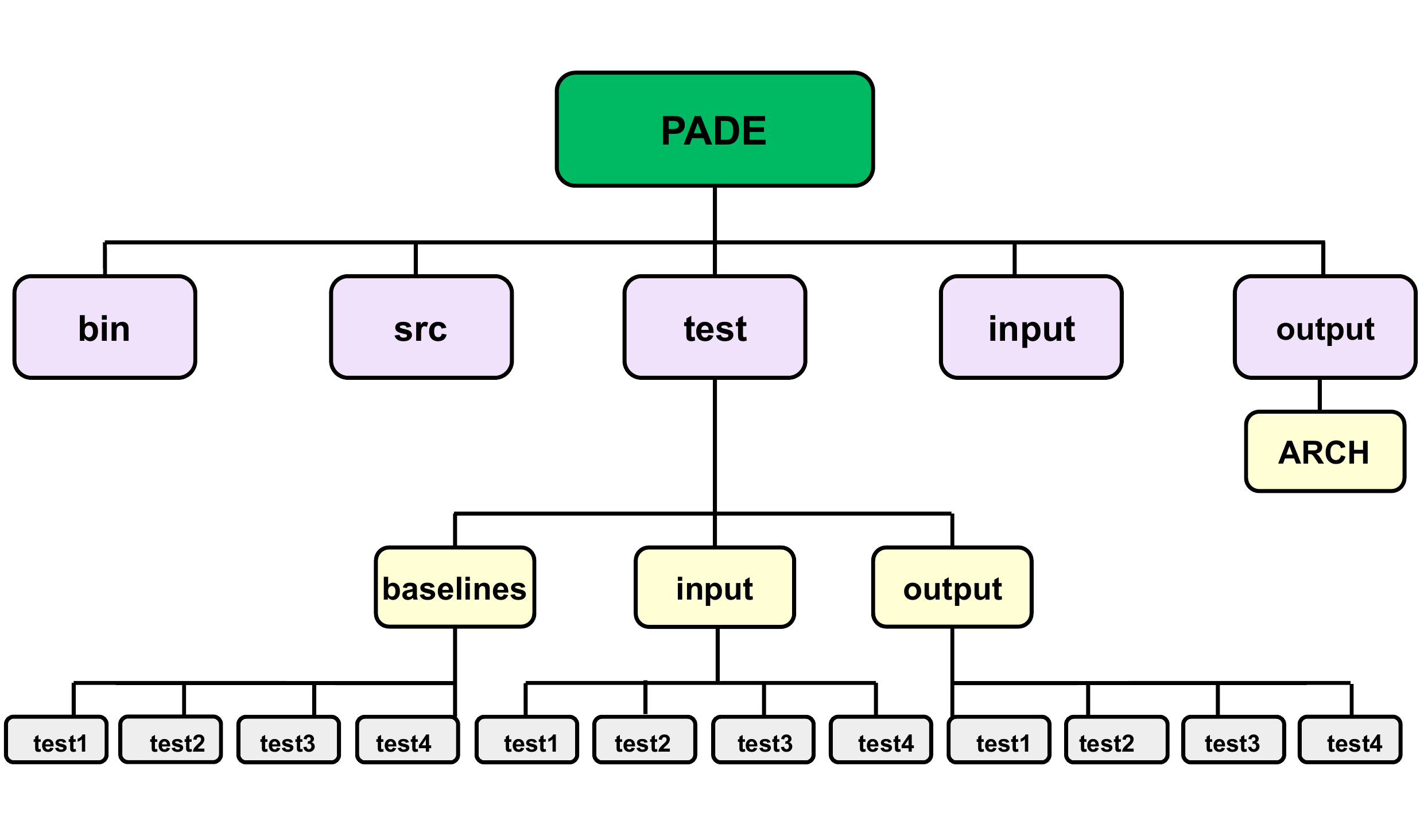}}
\caption{Detailed structure of the  \texttt{PADE} directory.}\label{plot:Fig14}
\end{figure}
See \autoref{plot:Fig14}.
\section{Appendix E: Structure of the \texttt{QUADPACK} directory}
\begin{figure}[ht]
 \subfloat{\includegraphics[angle=0,width=15 cm]{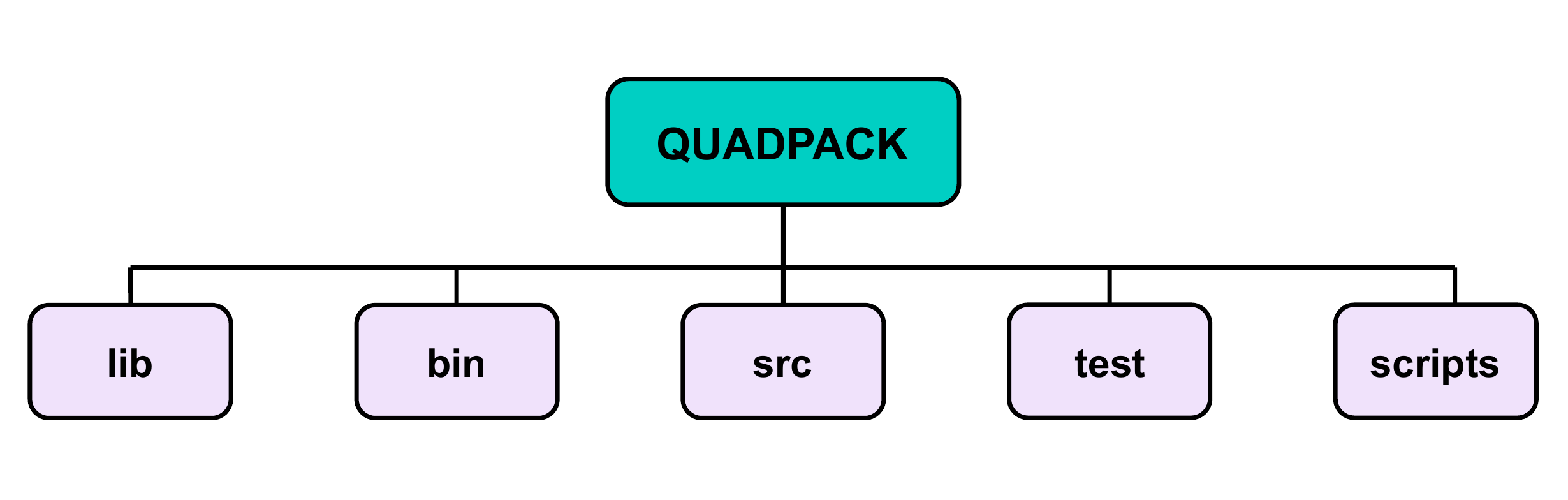}}
\caption{Detailed structure of the  \texttt{QUADPACK} directory.}\label{plot:Fig15}
\end{figure}
See \autoref{plot:Fig15}.
\newpage


\end{document}